\documentclass[conference]{IEEEtran}
\newif\ifsubmission

\usepackage{adjustbox}
\usepackage{amsmath,amssymb,amsfonts}
\usepackage[ruled,linesnumbered]{algorithm2e}
\usepackage{balance}
\usepackage{enumitem}
\usepackage[font={footnotesize}]{caption}
\usepackage{cite}
\usepackage{color, colortbl}
\usepackage{enumitem}
\usepackage{graphicx}
\usepackage[breaklinks,hidelinks]{hyperref}
\usepackage{cleveref} %

\usepackage{listings}
\usepackage{mdframed}
\usepackage{multirow}
\usepackage[numbers,sort]{natbib}
  \def\BibTeX{{\rm B\kern-.05em{\sc i\kern-.025em b}\kern-.08em
    T\kern-.1667em\lower.7ex\hbox{E}\kern-.125emX}}
\usepackage{pdfcomment}
\usepackage{relsize}
\usepackage{soul}
\usepackage{tabularx}
\usepackage{textcomp}
\usepackage{wasysym}
\usepackage{xcolor}
\usepackage{xspace}
\usepackage{float}
\usepackage{stmaryrd}
\usepackage{mathtools}
\usepackage{array}

\usepackage{anyfontsize}
\usepackage{graphicx}
\usepackage{soul}
\usepackage{array}
\usepackage{url}
\usepackage{multirow}
\usepackage{booktabs,makecell}
\usepackage{adjustbox}
\usepackage{array}
\usepackage{color}
\usepackage{amsmath}
\usepackage{bm}
\usepackage{url}
\usepackage{subcaption}
\usepackage{booktabs}
\usepackage{wasysym}
\usepackage{mathtools}
\usepackage{bm}

\usepackage{tikz}
\usetikzlibrary{arrows}

\usepackage{booktabs,subcaption,amsfonts,dcolumn}
\newcolumntype{P}[1]{>{\centering\arraybackslash}p{#1}}

\newcommand{\az}{\text{AndroZoo}\xspace}

\newcommand{\cpreject}{\text{CP-Reject}\xspace}
\newcommand{\ember}{\text{EMBER}\xspace}
\newcommand{\droidevolver}{\text{DroidEvolver}\xspace}
\newcommand{\hidost}{\text{Hidost}\xspace}

\newcommand{\eg}{e.g.,\xspace} %
\newcommand{\ie}{i.e.,\xspace} %
\newcommand{\cf}{cf.\xspace} %

\newcommand{\revised}[1]{#1}%

\newcommand{\algrule}[1][.2pt]{\par\vskip.2\baselineskip\hrule height #1\par\vskip.2\baselineskip}
\let\oldnl\nl%
\newcommand{\nonl}{\renewcommand{\nl}{\let\nl\oldnl}}%

\newcolumntype{R}[2]{%
    >{\adjustbox{angle=#1,lap=\width-(#2)}\bgroup}%
    c%
    <{\egroup}%
    |
}

\SetKwComment{Comment}{$\triangleright$\ }{}
\SetCommentSty{grey}

\DeclareMathOperator*{\argmax}{arg\,max}

\newcommand{\tool}{\textsc{Transcendent}\xspace}%
\newcommand{\transcendnocite}{\text{Transcend}\xspace}%
\newcommand{\transcend}{\text{Transcend}~\citep{jordaney2017transcend}\xspace}%
\newcommand{\tesseract}{\text{Tesseract}\xspace}%
\newcommand\tinybc{\vcenter{\hbox{\scalebox{0.5}{$\CIRCLE$}}}}%
\newcommand\tinywc{\vcenter{\hbox{\scalebox{0.5}{$\Circle$}}}}%
\newcommand{\Test}[1]{\textbf{\varhexstar}\xspace}%
\newcommand{\fone}{$F_1$\xspace}%
\newcommand{\miniclasscardinality}{\scalebox{0.5}{\ensuremath{|\mathcal{Y}|}}\xspace}
\newcommand{\uniformsample}{\xleftarrow{\scalebox{0.6}{\$}}\xspace}

\newcommand{\rulesep}{\textcolor{gray}{\unskip\ \vrule depth 1ex\ }}

\renewcommand{\paragraph}[1]{{\vskip 6pt \noindent\textbf{#1.} }}

\AtBeginDocument{%
	\Crefformat{section}{\S#2#1#3}%
	\crefformat{section}{\S#2#1#3}%
	\Crefformat{subsection}{\S#2#1#3}%
	\crefformat{subsection}{\S#2#1#3}%
	\Crefformat{subsubsection}{\S#2#1#3}%
	\crefformat{subsubsection}{\S#2#1#3}%
	\Crefrangeformat{section}{\S\S#3#1#4 to~#5#2#6}%
	\crefrangeformat{section}{\S\S#3#1#4 to~#5#2#6}%
	\Crefmultiformat{section}{\S\S#2#1#3}{ and~#2#1#3}{, #2#1#3}{ and~#2#1#3}%
	\crefmultiformat{section}{\S\S#2#1#3}{ and~#2#1#3}{, #2#1#3}{ and~#2#1#3}%
}

\IEEEoverridecommandlockouts
\begin{document}

\date{}

\title{Transcending {\sc Transcend}: Revisiting Malware Classification in the Presence of Concept Drift}

\ifsubmission
\author{ %
	--- Anonymous Author(s) ---
}
\else

\author{
    \IEEEauthorblockN{
    Federico~Barbero\thanks{\IEEEauthorrefmark{1}Equal contribution.}\IEEEauthorrefmark{1}\IEEEauthorrefmark{2}\IEEEauthorrefmark{6}, 
    Feargus~Pendlebury\IEEEauthorrefmark{1}\IEEEauthorrefmark{3}\IEEEauthorrefmark{4}\IEEEauthorrefmark{5}, 
    Fabio~Pierazzi\IEEEauthorrefmark{2}, 
    Lorenzo~Cavallaro\IEEEauthorrefmark{5}}
    
    \IEEEauthorblockA{
    \IEEEauthorrefmark{2} King's College London, 
    \IEEEauthorrefmark{3} Royal Holloway, University of London, 
    \IEEEauthorrefmark{4} The Alan Turing Institute, \\ 
    \IEEEauthorrefmark{6} University of Cambridge, 
    \IEEEauthorrefmark{5} University College London}
}

\fi

\maketitle
\ifsubmission
  \thispagestyle{plain}
  \pagestyle{plain}
\else
\fi

\begin{abstract}
  Machine learning for malware classification shows encouraging results,
  but real deployments suffer from performance degradation as malware
  authors adapt their techniques to evade detection. This phenomenon,
  known as \textit{concept drift}, occurs as
  new malware examples evolve and become less and less like the original
  training examples.
  One promising method to cope with concept drift is \textit{classification with rejection} in which examples that are likely to be misclassified are instead quarantined until they can be expertly analyzed.

We propose \tool, a rejection framework built on \transcendnocite, a recently proposed strategy based on conformal prediction theory. In particular, we
provide a formal treatment of \transcendnocite, enabling us to refine
\textit{conformal evaluation theory}---its underlying statistical
engine---and gain a better understanding of the theoretical reasons for
its effectiveness.  In the process, we develop two additional conformal
evaluators that match or surpass the performance of the original while
significantly decreasing the computational overhead. We evaluate \tool on a malware dataset spanning 5 years that removes sources of experimental bias
present in the original evaluation. \tool outperforms
state-of-the-art approaches while generalizing across different malware
domains and classifiers.

To further assist practitioners, we showcase optimal operational
settings for a \tool deployment and show how it can be applied to
many popular learning algorithms. These insights support both old and new empirical findings, making \transcendnocite a sound and practical solution for the first time. To this end, we release \tool as open source, to aid the adoption of rejection strategies by the security community.

\end{abstract}


\section{Introduction}
\label{sec:intro}

Machine learning (ML) algorithms have displayed superhuman performance across a wide range of classification tasks such as computer vision~\citep{alexnet} and natural language processing~\citep{backtranslate}. However, a great deal of this success is conditional on one central assumption: that the training and test data are drawn identically and independently from the same underlying distribution (i.i.d.)~\cite{bishop}.

In a security setting this assumption often does not hold. In particular, malware classifiers are deployed in dynamic, hostile environments~\citep{pendlebury2021hostile}. New paradigms of malware evolve to pursue profits, new variants arise as novel exploits are discovered, and adversaries switch behavior suddenly and dramatically when faced with strengthened defenses. This causes the incoming test distribution to diverge from the original training distribution, a phenomenon known as  \textit{concept drift}~\citep{datasetshift}. Over time, classifier performance begins to degrade as the model fails to classify the new objects correctly.

There appear to be two broad approaches to tackling concept drift. The
first is to design systems which are intrinsically more \emph{resilient} to
drift by developing more robust feature spaces. For example, it has
recently been suggested that neural networks may be more resilient to
concept drift as the latent feature space better generalizes to new
variants~\cite{pendlebury2019tesseract}. However, designing robust
feature spaces is an open research question and it is not clear if there
exists such a malware representation for which concept drift will
not occur.

A second solution is to \textit{adapt} to the drift, for example by updating the model using incremental retraining or online learning~\citep{xu2019edvolver, casandra}, or rejecting drifting points. However, to be effective, decisions about when and how to take action on aging classifiers must be taken quickly and decisively. To do so, accurate detection and quantification of drift is vital.

This problem is precisely the focus of
\transcend, a statistical framework that
builds on conformal prediction theory~\cite{randomworld} to detect aging
malware detectors during deployment---before their accuracy deteriorates
to unacceptable levels.
\transcend proposes a \textit{conformal
evaluator} that utilizes the notion of \textit{nonconformity} to
identify and reject new examples that differ from the training
distribution and are likely to be misclassified; the corresponding apps
can then be quarantined for further analysis and labeling. While
effective, the original proposal suffers from experimental bias, is
extremely resource intensive and thus impractical, lacks experiments to support generalization claims, fails to provide
guidance on how to integrate it into a detection pipeline and, perhaps
more importantly, lacks a theoretical analysis to explain
its effectiveness.

In this paper, we revisit conformal evaluator and
\transcendnocite to root its internal workings
in sound theory and determine its most effective operational settings. We additionally propose \tool, a framework that surpasses the
performance of the original in terms of drift detection and
computational overhead, making it a sound and practical solution.

In summary, we make the following contributions:

\begin{itemize}
	\item \textbf{Formal Treatment.} We investigate the theory underpinning the motivation and intuition of conformal evaluation to provide a missing link between conformal evaluation and conformal prediction theory that explains its effectiveness and supports the empirical evaluations presented in both this work and the original~(\cref{sec:ce}).
	\item \textbf{Novel Conformal Evaluators.} Building on this insight, we develop two novel conformal evaluators: \textit{inductive conformal evaluator} (ICE)~(\cref{sec:ice}) and \textit{cross-conformal evaluator} (CCE)~(\cref{sec:cce}), both of which are firmly grounded in conformal prediction theory and able to effectively identify and reject drifting examples while being significantly less computationally demanding than the original. We formalize \transcendnocite's calibration procedure as an optimization problem and propose an improved search strategy for finding thresholds (\cref{sec:transcendframework}).
	\item \textbf{Operational Guidance.} We evaluate our proposals on a dataset spanning 5 years (2014--2019)
	that eliminates sources of bias present in past evaluations~(\cref{sec:eval}). We compare different operational settings, including the effects of including algorithm \textit{confidence}~(\cref{sec:credconfeval}) and of using different search strategies~(\cref{sec:search-eval}) during thresholding. \revised{Our methods outperform existing state-of-the-art approaches~(\Cref{sec:prior-methods}), and generalize well across different malware domains and underlying classifiers~(\Cref{sec:beyond-android}).} To aid practitioners in adopting rejection strategies, we include a discussion of how to integrate \tool into a typical security detection pipeline~(\cref{sec:discussion}).
\end{itemize}

To enable researchers and practitioners to make better use of classification with rejection strategies, we publicly release our data and implementation of \tool.\footnote{\url{https://s2lab.cs.ucl.ac.uk/projects/transcend/}}%

\section{Background}
\label{sec:background}

We focus on classification for security tasks~(\Cref{sec:bg-mlsec}) which are affected by concept drift~(\Cref{sec:bg-drift}). In particular, we are interested in improving the state-of-the-art approaches for classification with rejection~(\Cref{sec:bg-rejection}).

\subsection{Machine Learning and Security Detection}
\label{sec:bg-mlsec}

Machine learning is a set of statistical methods for automating data analysis and enabling systems to perform tasks on the data without being explicitly programmed for them. In the malware domain, typical tasks include binary classification (detecting malicious examples~\citep{arp2014drebin, xu2019edvolver}) and multiclass classification (predicting the malware family~\citep{guillermo2014dendroid,dash2016droidscribe,guillermo2017droidsieve}) but can also extend to more complex tasks such as predicting how many AV engines would detect an example~\citep{kantchelian2015regression}, inferring Android malware app permissions based on their icons~\citep{xi2019deepintent}, or generating Windows malware using reinforcement learning~\citep{anderson2018pdfgan}.

In this paper we focus on classification tasks where a classifier $g$ aims to learn a function mapping $\mathcal{X} \rightarrow \mathcal{Y}$, where $\mathcal{X} \subseteq \mathbb{R}^n$ is a feature space of vectors capturing interesting properties of the apps and $\mathcal{Y}$ is a label space containing binary labels for the detection task or \revised{the names of malware families for the multiclass classification task}.

\subsection{Concept Drift}
\label{sec:bg-drift}

One of the greatest challenges facing machine learning-based malware classifiers is the presence of \textit{dataset shift}~\cite{miller16dimva,allix15relevant,jordaney2017transcend} as the distribution of malware at test time begins to diverge from the training distribution. This violates one of the core assumptions of most classification algorithms: that the training and test time examples are identically and independently drawn from the same joint distribution (i.i.d.). As this assumption weakens over time, the classifier's predictions become less and less reliable and performance degrades.

Dataset shift can be broadly categorised into three types of shift~\cite{datasetshift}. \textit{Covariate shift} refers to a change in the distribution of $P(\bm{x} \in \mathcal{X})$, when the frequency of certain features rises or falls (\eg variations in API call frequencies over time). \textit{Prior probability shift} or \textit{label shift} is a change in the distribution of $P(y \in \mathcal{Y})$, when the base rate of a particular class is altered (\eg an increase in malware prevalence over time). \textit{Concept drift} is a change in the distribution $P(y \in \mathcal{Y} | \bm{x} \in \mathcal{X})$. This often occurs when the definition of the ground truth changes, \eg, if a new family of malware arises which, given the feature space representation $\mathcal{X}$, is indistinguishable from benign applications. Due to limited knowledge, the model will start misclassifying examples from the new family, even if no covariate or prior probability shift has occurred. In practice, it can be extremely difficult to determine how much error should be attributed to each type of shift~\cite{datasetshift}. Given this, it is common in the security community to collectively refer to all types of shift as \textit{concept drift}, a custom that we continue in this work.

The impetus for concept drift in malware classification is the adversarial nature of the task. Malware authors are driven by the profit motive to try and evade detection or classification by app store owners, antivirus companies, and users. This incentivizes them to innovate: to obfuscate features of their malware, develop new methods for exploitation and persistence, and explore new avenues of profiteering and abuse. This causes the definition of malware to evolve over time, sometimes in drastic or unexpected ways.

\begin{figure*}[t!]
    \centering
        \begin{subfigure}[t]{0.21\textwidth}
        \centering
        \includegraphics[width=\textwidth]{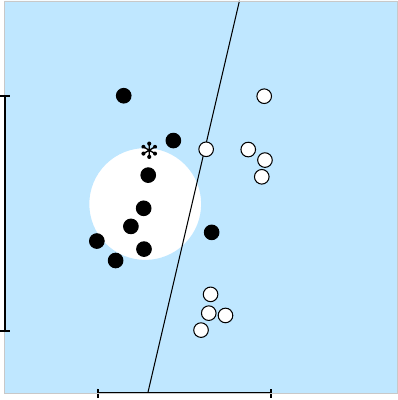}
        \caption{Nearest centroid}
        \label{fig:ncm-centroid}
    \end{subfigure}
    \hfill
    \begin{subfigure}[t]{0.21\textwidth}
        \centering
        \includegraphics[width=\textwidth]{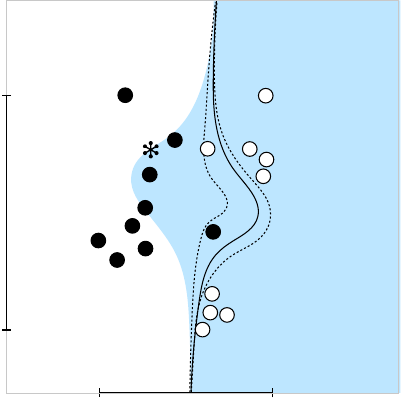}
        \caption{Polynomial SVM}
        \label{fig:ncm-poly}
    \end{subfigure}%
    \hfill
    \begin{subfigure}[t]{0.21\textwidth}
        \centering
        \includegraphics[width=\textwidth]{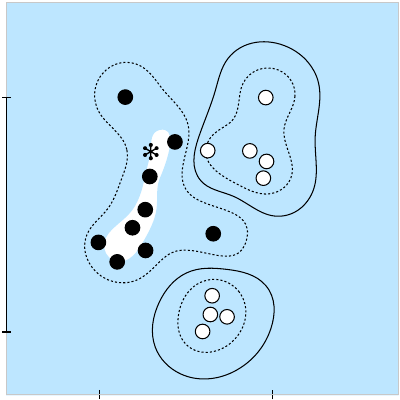}
        \caption{RBF SVM}
        \label{fig:ncm-rbf}
    \end{subfigure}%
    \hfill
    \begin{subfigure}[t]{0.21\textwidth}
        \centering
        \includegraphics[width=\textwidth]{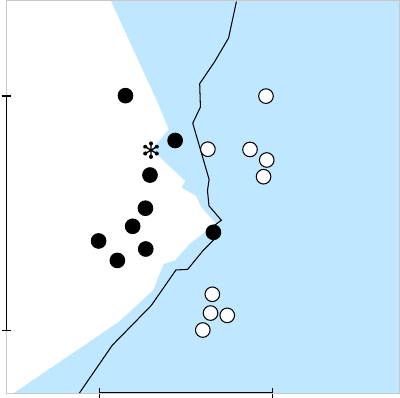}
        \caption{3-NN}
        \label{fig:ncm-knn}

    \vspace{0.5em}
    \end{subfigure}%

    \begin{subfigure}[t]{0.21\textwidth}
        \centering
        \includegraphics[width=\textwidth]{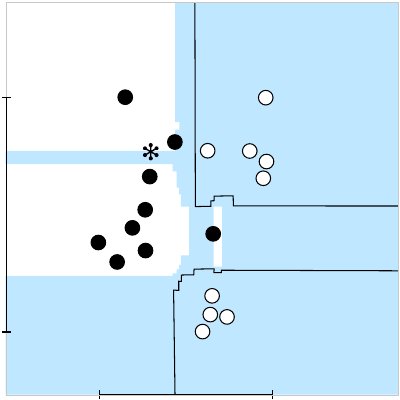}
        \caption{Random forest}
        \label{fig:ncm-rf}
    \end{subfigure}%
    \hfill
    \begin{subfigure}[t]{0.21\textwidth}
        \centering
        \includegraphics[width=\textwidth]{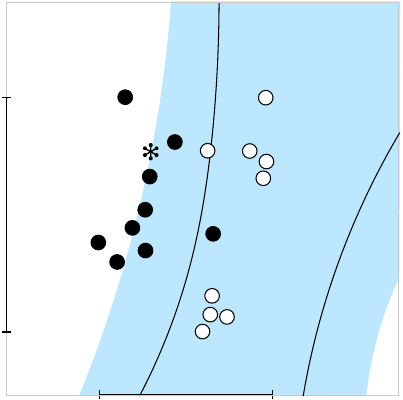}
        \caption{QDA}
        \label{fig:ncm-qda}
    \end{subfigure}%
    \hfill
    \begin{subfigure}[t]{0.21\textwidth}
        \centering
        \includegraphics[width=\textwidth]{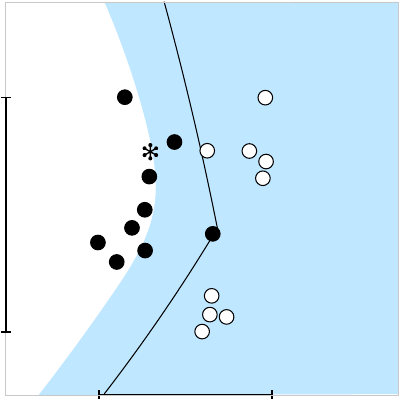}
        \caption{MLP sigmoid}
        \label{fig:ncm-nn}
    \end{subfigure}%
    \hfill
    \begin{subfigure}[t]{0.21\textwidth}
        \centering
        \includegraphics[width=\textwidth]{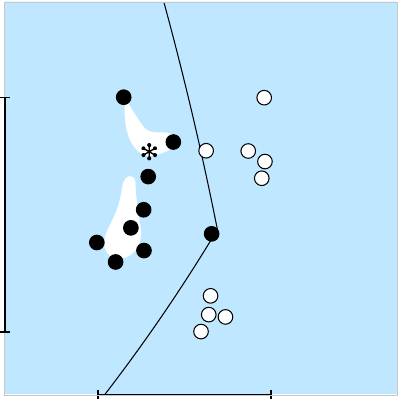}
        \caption{MLP with SVM RBF}
        \label{fig:ncm-nnsvm}
    \end{subfigure}
    \caption{Possible NCMs for different classification algorithms: nearest centroid, support-vector machines (SVMs), nearest neighbors (NN), random forest, quadratic discriminant analysis (QDA), and multilayer perceptron (MLP). The solid line delineates the  decision boundary between classes \CIRCLE{} and \Circle{} while the dotted lines show SVM margins. The shaded region captures points which are \textit{more nonconform} (\ie `less similar') than the new test point, shown by the asterisk, with respect to class \CIRCLE{}. As NCMs, (a) uses the distance from the class centroid; (b) and (c) use the negated absolute distance from the hyperplane; (d) uses the proportion of nearest neighbors belonging to class \Circle{}; (e) uses the proportion of decision trees that predict \Circle{}; (f) uses the negated probability of belonging to class \CIRCLE{}; (g) uses the negated probability output by the final sigmoid activation layer; (h) uses the outputs of the final hidden layer to train an SVM with RBF kernel and uses the negated absolute probabilities output by that SVM---note the decision boundary still depends on the MLP output alone.}
    \label{fig:ncm-examples}
\end{figure*}

\subsection{Rejection}
\label{sec:bg-rejection}

There are multiple routes to dealing with concept drift. The most effective would be to design a feature space $\mathcal{X}$ such that it is entirely robust to concept drift, essentially distilling all possible malware behaviour down to a `Platonic ideal'~\cite{plato} that captures maliciousness no matter what form it takes. While recent proposals for augmenting feature spaces with robust features are promising~\citep[\eg][]{liang2019conserved,zhang2020apigraph}, the diversity of malware makes it extremely difficult to design such a feature space. Additionally, some behaviour is only considered malicious due to its context, for example, requesting access to the device contacts might be considered suspicious for a torch app but not for a social messaging app~\cite{yang2015appcontext}.

An orthogonal approach is to identify, track, and mitigate the drift as it occurs. One promising method is classification with rejection~\cite{bartlett08rejection}, in which low confidence predictions, caused by drifting examples, are \textit{rejected}. Drifting apps can then be quarantined and dealt with separately, either warranting manual inspection or remediation through other means.

\transcend is a state-of-the-art framework for performing classification with rejection in security tasks. It uses a \textit{conformal evaluator} to generate a quality measure to assess whether a new test example is drifting with respect to the training data. If the prediction of an underlying classifier appears to be affected by the drift, the prediction is rejected. The original proposal presented two case studies: Android malware detection---a binary classification task, and Windows malware family classification---a multiclass classification task. The experiments showed that the framework is consistently able to identify drifting examples, providing a significant improvement over thresholding on the classifiers' output probabilities. However, the lack of a theoretical treatment and the computational complexity of the framework limited its understanding and use in real-world deployments.

\section{Towards Sound Conformal Evaluation}
\label{sec:ce}

The statistical engine that drives \transcendnocite's rejection mechanism is the \textit{conformal evaluator}, a tool for measuring the quality of predictions output by an underlying classifier. Conformal evaluator design is grounded in the theory of \textit{conformal prediction}~\cite{randomworld}, a method for providing predictions that are correct with some guaranteed confidence. In this section we investigate the relationship between the two to provide novel insights and intuition into why conformal evaluation is effective in the classification with rejection setting.

\subsection{Conformal Evaluation vs. Prediction}
\label{sss:cp}
Here we give an overview of conformal prediction and how it motivates the use of conformal evaluation; for a more formal treatment of conformal prediction we refer to \citet{randomworld}.
Conformal prediction allows for predictions to be made with precise levels of confidence by using past experience to account for uncertainty.
Given a classifier $g$, a new example $z = (x, y)$, and a significance level $\varepsilon$, a conformal predictor produces a \textit{prediction region}: a set of labels in the label space $\mathcal{Y}$ that is guaranteed to contain the correct label $y$ with probability no more than $1 - \varepsilon$. To calculate this label set, the conformal predictor relies on a \textit{nonconformity measure} (NCM) derived from $g$ and uses it to generate scores representing how \textit{dissimilar} each example is from previous examples of each class. To quantify this relative dissimilarity, \textit{p-values} are calculated by comparing the nonconformity scores between examples~(\cref{sss:ncms}). As well as these p-values, two important metrics are derived from the prediction region, confidence and credibility~(\cref{sss:credconf}), which can be used to judge the effectiveness of the conformal prediction framework. Conformal predictors are able to make strong guarantees on the correctness of each prediction so long as two assumptions about new test examples hold: the \textit{exchangeability} assumption, that the sequence of examples is exchangeable, a generalization of the i.i.d. property; and the \textit{closed-world} assumption, that new examples belong to one of the classes observed during training.

Rather than making predictions, conformal evaluators~\cite{jordaney2017transcend} borrow the same statistical tools (\ie nonconformity measures and p-values) but use them to \textit{evaluate} the quality of the prediction made by the underlying classifier $g$. By detecting instances which appear to violate the aforementioned assumptions they can, with high confidence, \textit{reject} new drifting examples which would otherwise be at risk of being misclassified.

\subsection{Nonconformity Measures and P-values}
\label{sss:ncms}

In order to reject a new example that cannot be reliably classified, conformal evaluators rely on a notion of \textit{nonconformity} to quantify how dissimilar the new example is to a history of past examples.
In general, a \textit{nonconformity measure}~(NCM)~\cite{shafer2008tutorial} is a real-valued function that outputs a score describing how different an example $z$ is from a bag of previous examples $B = \Lbag z_1, z_2, ... , z_n \Rbag$:
\begin{align}
	\alpha_z = A(B, z).
\end{align}

The greater the value of $\alpha_z$, the less similar $z$ is to the elements of the bag $B$. An NCM is typically formed of two components: a metric $d(z, z')$ to measure the distance between two points, and a \textit{point predictor} $\hat{z}(B)$ to represent $B$:
\begin{align}
	A(B, z) \coloneqq d(\hat{z}(B), z). \label{eq:ncm-form}
\end{align}

Illustrating this, \autoref{fig:ncm-centroid} shows an NCM for a nearest centroid classifier in which the Euclidean distance is used for $d(z, z')$, and the nearest class centroid is used for $\hat{z}(B)$.

For a new example $z^*$, the conformal evaluator must decide whether or not to approve the null hypothesis asserting that $z^*$ \textit{does not belong} in the prediction region formed by elements of $B$. To perform such a hypothesis test, \textit{p-values} are calculated using the NCM values for each point. First the nonconformity score of $z^*$ must be computed~(\autoref{eq:test-ncm}) along with nonconformity scores of elements in $B$~(\autoref{eq:train-ncms}), then the the p-value $p_{z^*}$ for $z^*$ is given as the proportion of points with greater or equal nonconformity scores~(\autoref{eq:test-pval}):
\begin{align}
	\alpha_{z^*} &= A(B, z^*) \label{eq:test-ncm} \\
	S &= \Lbag A(B \setminus \Lbag z \Rbag, z) : z \in B \Rbag \label{eq:train-ncms}\\
	p_{z^*} &= \frac{| \alpha \in S : \alpha >= \alpha_{z^*} |}{|S|}
	\label{eq:test-pval}
\end{align}
In the classification context, we can calculate p-values in a \textit{label conditional} manner, such that $B$ contains only previous examples of class $\hat{y} \in Y$ where $\hat{y} = g(z^*)$ is the predicted class of the new example. If $p_{z^*}$ falls \revised{above} a given significance level the null hypothesis is disproved and $\hat{y}$ is accepted as a valid prediction. \transcend computes \textit{per-class thresholds} to use as significance levels~(\cref{sec:transcendframework}).

As p-values are calculated by considering nonconformity scores relative to one another, NCMs can be transformed monotonically without any impact on the resulting p-values. Thus, when designing an NCM in the form given by \autoref{eq:ncm-form}, the distance metric $d(z, z')$ is significantly less important than the point predictor $\hat{z}(B)$. It is important to note that conformal evaluator algorithms are agnostic to the underlying NCM chosen, but the quality of the NCM---and particularly of $\hat{z}(B)$, will impact the ability of conformal evaluators to discriminate between valid and invalid predictions~\cite{shafer2008tutorial}.

An \textit{alpha assessment}~\cite{jordaney2017transcend} can be used to empirically evaluate how appropriate an NCM is for a given dataset by plotting the distribution of p-values for each class, further split into whether the prediction was correct or incorrect. As incorrect predictions should be rejected, they are expected to fall below the threshold, while correct predictions are expected to fall above the threshold. Well-separated distributions of correct and incorrect predictions suggest a viable threshold exists to separate them at test time. Poorly separated prediction p-values indicate an inappropriate NCM. An example of an alpha assessment on a toy dataset is shown in \autoref{fig:thresh-alpha}.

\autoref{fig:ncm-examples} illustrates possible NCMs for different algorithms on a toy binary classification task with existing class examples~\CIRCLE{}/\Circle{} and new test example \Test{}. The solid line delineates the decision boundary between the two classes, the dotted lines show SVM margins where applicable, and the blue shaded region captures points that are \textit{more nonconform} (\ie less similar) than \Test{} with respect to class~\CIRCLE{}. Note that the shape of the nonconformal region need not reflect the shape of the regions for the predicted classes (\eg \autoref{fig:ncm-centroid}) and that there may be multiple viable NCMs for the same underlying algorithm (\eg \Cref{fig:ncm-nn,fig:ncm-nnsvm}).

\begin{figure}[t!]
    \centering
	\includegraphics[width=0.99\columnwidth]{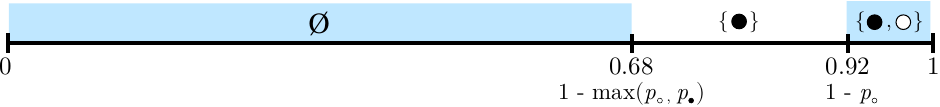}
    \caption{The nested intervals at which labels \CIRCLE{} and \Circle{} are present in the output label set for a test example with per-class p-values $p_{\protect\tinybc}=0.32$ and $p_{\protect\tinywc}=0.08$. Shaded areas outline how credibility and confidence relate to the intersection of prediction regions for which the label set contains a single element. The relatively high probability of the empty set containing the correct label (\ie low credibility) indicates that one of conformal prediction's assumptions may have been violated. In conformal evaluation, this is used as a signal that the new example is likely  out-of-distribution and is indicative of concept drift.}
    \label{fig:cp-prediction-region}
\end{figure}

\subsection{Successfully Identifying Drift}
\label{sss:credconf}

Recall that conformal prediction produces a prediction region given a significance level $\varepsilon$. The possible prediction regions are nested such that the higher the confidence level, the more labels will be present. As a trivial example, a prediction region containing all possible labels may be produced for a significance level of $\varepsilon = 0$ (maximum likelihood) as it will contain the true label $y$ with certainty. At the other extreme, an empty set can be produced at a significance level of $\varepsilon = 1$ (minimum likelihood), as this is an impossible result under the closed-world assumption of conformal prediction.

Of particular interest is the prediction region containing a single element which lies between these extremes.
Related to this prediction region, a conformal predictor also outputs two metrics: \textit{confidence} and \textit{credibility} (\autoref{fig:cp-prediction-region}).

Confidence is the greatest $1 - \varepsilon$ for which the prediction region contains a single label which can be calculated as the complement to 1 of the second highest computed p-value. Confidence quantifies the likelihood that the new element belongs to the predicted class.

Credibility is the greatest $\varepsilon$ for which the prediction region is empty and corresponds to the largest computed p-value. Conformal predictors can be forced to output single predictions (rather than a label set induced by $\varepsilon$), for which they will output the class with the highest credibility. Credibility quantifies how relevant the training set is to the prediction. A low credibility indicates conformal prediction might not be a suitable framework to use with the given data because a low credibility means the probability of the correct label being in the empty set is relatively high, which is an impossible result under the closed-world assumption of conformal prediction.

We propose that conformal evaluation's effectiveness stems from this relationship: that in conformal \textit{evaluation}, this probability is being directly interpreted as the probability that the i.i.d. assumption has been violated. Thus, a low credibility means that there is a high probability that the corresponding example is \textit{drifting} with respect to the previous history of training examples. Such an example is at risk of being misclassified due to limited knowledge of the classifier.

It should be noted that formally, conformal evaluation defines credibility and confidence slightly differently. In conformal evaluation, the credibility is the p-value corresponding to the predicted class and the confidence is the complement to 1 of the maximum p-value excluding the p-value corresponding to the predicted class (\ie the credibility p-value). This subtle difference is important to clarify the operational context of a conformal evaluator: whereas conformal predictors output the final classification decision, conformal evaluators output a statistical measure \textit{separate} to the decision of the underlying classifier (hence the nomenclature: one predicts and the other evaluates). In practice, given reasonable NCMs, these definitions can be treated as equivalent.

\section{Towards Practical Conformal Evaluation}
\label{sec:novel-ce}

In assessing the quality of a prediction for a new test point, there is the question of which previously encountered points the new point should be compared to---that is, which elements are included in the bag $B$ of \autoref{eq:test-ncm}, and how. Typically, new test points are compared against a set of calibration points.

\label{sec:tce}

In \citet{jordaney2017transcend}, conformal evaluation was realized using a Transductive Conformal Evaluator (TCE). With a TCE, every training point is also used as a calibration point. To generate the p-value of a calibration point, it is first removed from the set of training points and the underlying classifier trained on the remaining points. Given the newly trained classifier, a predicted label is generated for the calibration point. Finally, using a given NCM, its p-value is computed with respect to the points whose ground truth label matches its predicted label. This procedure is repeated for every training point. Following this, \transcendnocite's thresholding mechanism operates on the calculated p-values to determine per-class rejection thresholds~(\cref{sec:transcendframework}). At test time, the underlying classifier is retrained on the entire training set, and, similarly to the calibration points, the p-values are computed with respect to the p-values of the calibration sets.

While the Transductive Conformal Evaluator (TCE) used in the original proposal~\cite{jordaney2017transcend} appears to perform well, \revised{it does not scale to larger datasets as a newly trained classifier is required for every training point.}
Consider the experiments in~\cref{sec:eval} where fitting a single instance of the underlying classifier takes 10 CPU minutes. In this case, we estimate a single run using vanilla TCE to take 1.9 CPU years.

\begin{figure}[t!]
    \centering

    \begin{subfigure}[t]{0.99\columnwidth}
        \centering
        \includegraphics[height=0.5in]{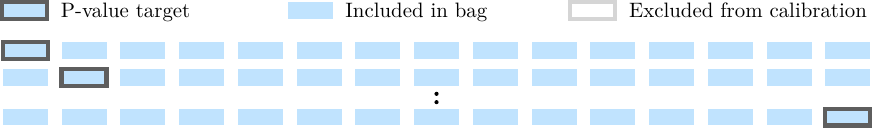}
        \caption{TCE}
        \label{fig:splits-tce}
    \vspace{0.5em}
    \end{subfigure}

    \begin{subfigure}[t]{0.99\columnwidth}
        \centering
        \includegraphics[height=0.35in]{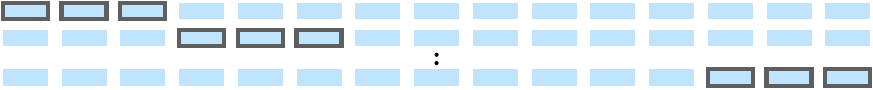}
        \caption{Approx-TCE}
        \label{fig:splits-approx-tce}
    \vspace{0.5em}
    \end{subfigure}%

    \begin{subfigure}[t]{0.99\columnwidth}
        \centering
        \includegraphics[height=0.35in]{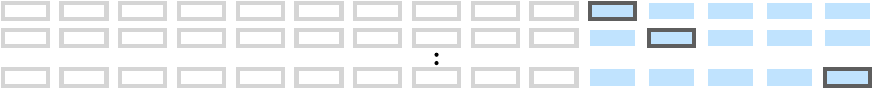}
        \caption{ICE}
        \label{fig:splits-ice}
    \vspace{0.5em}
    \end{subfigure}%

    \begin{subfigure}[t]{0.99\columnwidth}
        \centering
        \includegraphics[height=1.275in]{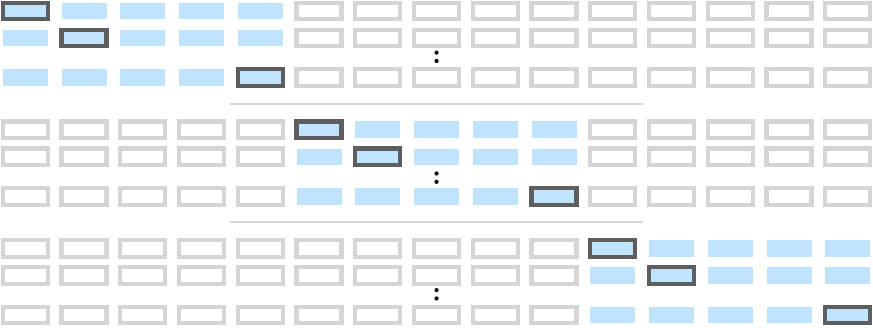}
        \caption{CCE}
        \label{fig:splits-cce}
    \end{subfigure}%

    \caption{Illustration of the different calibration splits employed by each of the conformal evaluators showing the target of the p-value calculation, relative points included in the bag, and points excluded from the calibration.}
    \label{fig:ce-splits}
\end{figure}

We propose a number of novel conformal evaluators that overcome this limitation and present their advantages and disadvantages. A comparison of their runtime complexities and operational considerations are presented in~\autoref{tab:complexities} and~\Cref{sec:discussion}, respectively. Formal algorithms for their calibration and test procedures are included in~\Cref{app:algos} while~\autoref{fig:ce-splits} provides a graphical intuition to their different calibration splits.

\revised{Note that while our illustrative examples and evaluation are given for the binary detection task, \tool and conformal evaluation are agnostic to the total number of classes and this is captured in the formal definitions. If multiclass NCMs cannot be derived, per-class conformal evaluators may be arranged as a \textit{one-vs-all} ensemble.}

\subsection{Approximate TCE (approx-TCE)}
\label{sec:approx-tce}

Our first attempt at reducing the computational overhead induced by the Transductive Conformal Evaluator is the \textit{approximate Transductive Conformal Evaluator} (approx-TCE). In the original TCE, p-values are generated for each calibration point by removing them from the training set, retraining the underlying classifier on the remaining points, and repeating until a p-value is computed for every training point.

In approx-TCE, calibration points are left out in batches, rather than individually. The training set is randomly partitioned into $k$ folds of equal size. From the $k$ folds, one is used as the target of the calibration and the remaining $k - 1$ folds are used as the bag to which those points are compared to. This process repeats $k$ times, until each fold has been used as the calibration set exactly once. Note that all of the $k$ calibration sets are mutually exclusive; the corresponding batches of p-values are then concatenated in the same manner as in TCE.

The statistical soundness of the approx-TCE relies on the assumption that the decision boundary obtained from leaving out calibration points in batches approximates each of the decision boundaries that would have been obtained per calibration point in the batch if the point had been left out individually. If this assumption holds, the generated p-values will be the same as, or similar to, the p-values generated with a TCE. The approximation grows more accurate as $k$ increases until $k$ equals the cardinality of the training set at which point the approx-TCE and the TCE are equivalent. In this sense, the approx-TCE can be viewed as a generalization of the TCE.

This assumption is more likely to hold with algorithms with lower variance (\eg linear models), but becomes more tenuous as the variance increases unless $k$ increases also---sacrificing the saved computation to mitigate the statistical instability.

\subsection{Inductive Conformal Evaluator (ICE)}
\label{sec:ice}

The second conformal evaluator we propose is the \textit{Inductive Conformal Evaluator} (ICE) which, unlike the approx-TCE, is based on a corresponding approach from conformal prediction theory~\cite{randomworld, vovk13inductive, papadopoulos08inductive}. The ICE directly splits the training set into two non-empty partitions: the \textit{proper training set} and the \textit{calibration set}. The underlying algorithm is trained on the proper training set, and p-values are computed for each example in the calibration set. Unlike the TCE, p-values are not calculated for every training point, but only for examples in the calibration set, with the proper training set having no role in the calibration at all. The ICE aims to inductively learn a \textit{general rule} on a single fold of the training set.

This induces significantly less computational overhead than the TCE and approx-TCE (see~\autoref{tab:complexities}) and in practice is extremely fast, but also very \textit{informationally inefficient}. Only a small proportion of the training data is used to calibrate the conformal evaluator, when ideally we would use all of it. Additionally, the performance of the evaluator depends heavily on the quality of the split and the calibration set's ability to generalize to the remainder of the dataset. This results in some uncertainty: an ICE may perform worse than a TCE due to a lack of information, or better due to a lucky split.

\subsection{Cross-Conformal Evaluator (CCE)}
\label{sec:cce}

The \textit{Cross-Conformal Evaluator} (CCE) draws on inspiration from $k$-fold cross validation and aims to reduce both the computational and informational inefficiencies of the TCE and ICE. Like the ICE, the CCE has a counterpart rooted in conformal prediction theory~\cite{vovk2018ccp}.

The training set is partitioned into $k$ folds of equal size. So that a p-value is obtained for every training example, each fold is treated as the calibration set in turn, with p-values calculated as with an ICE, using the union of the $k - 1$ remaining folds as the proper training set to fit the underlying classifier.

Finally we concatenate the p-values in a way which preserves their statistical integrity when decision quality is evaluated. We set aside the $k$ fit underlying models and corresponding calibration sets for test time. When a new point arrives, the prediction from each classifier is evaluated against the corresponding calibration set. The final result is the majority vote over the $k$ folds, \ie the prediction of a particular class is accepted if the number of accepted classifications is greater than $\frac{k}{2}$, and rejected otherwise.

The CCE can be viewed as $k$ ICEs, one per fold, and these ICEs can operate in parallel to reduce computation time---if the resources are available. However, there is an additional memory cost with storing the separate models.

\begin{table}[t!]
\centering
\footnotesize
\caption{Runtime complexities and empirical runtime for conformal evaluator calibration where $n$ is the number of training examples and $p$ is the proportion of examples included in the \textit{proper training set} each split/fold.}
\begin{tabular}{llr}
  \toprule
  {\sc Conformal Evaluator} & {\sc Complexity} & {\sc Runtime in \Cref{sec:conf-eval}} \\
  \midrule
    TCE  							& $\mathcal{O}(n^2)$          & \textcolor{gray}{est. $1.9$ CPU yrs} \\
    Approx-TCE, $1 / (1 - p)$ folds 	& $\mathcal{O}(n / (1 - p))$  & $46.1$ CPU hrs \\
    ICE  							& $\mathcal{O}(pn)$           & $11.5$ CPU hrs \\
    CCE, $1 / (1 - p)$ folds		& $\mathcal{O}(pn / (1 - p))$ & $36.6$ CPU hrs \\
  \bottomrule
\end{tabular}

\label{tab:complexities}
\end{table}

\begin{figure*}[t!]
    \centering
    \begin{subfigure}[t]{0.24\textwidth}
        \centering
        \includegraphics[height=1.3in]{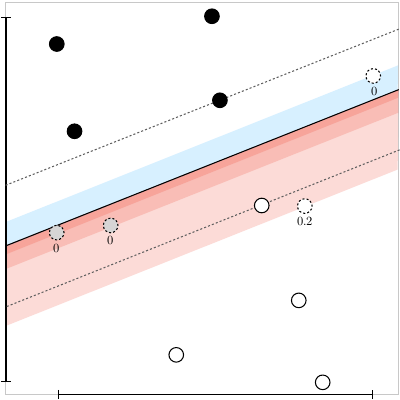}
        \caption{First fold}
        \label{fig:thresh-one}
    \end{subfigure}%
    ~
    \begin{subfigure}[t]{0.24\textwidth}
        \centering
        \includegraphics[height=1.3in]{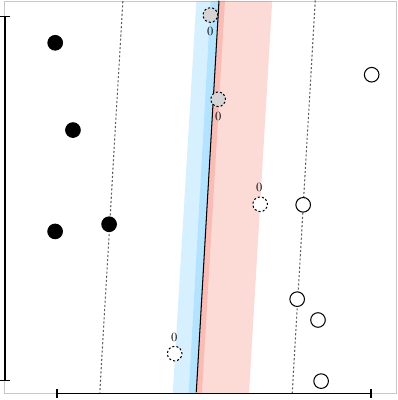}
        \caption{Second fold}
        \label{fig:thresh-two}
    \end{subfigure}%
    ~
    \begin{subfigure}[t]{0.24\textwidth}
        \centering
        \includegraphics[height=1.3in]{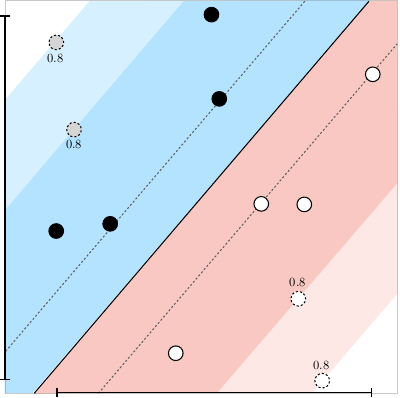}
        \caption{Third fold}
        \label{fig:thresh-three}
    \end{subfigure}%
    ~
    \begin{subfigure}[t]{0.24\textwidth}
        \centering
        \includegraphics[height=1.3in]{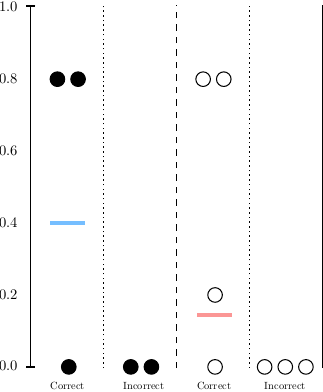}
        \caption{Alpha assessment}
        \label{fig:thresh-alpha}
    \end{subfigure}
    \caption{Thresholding procedure applied to a linear SVM with approximate-TCE (3 folds). Four points highlighted with dotted outlines are left out as calibration in each fold, with the decision boundary obtained with the remaining points as training. P-values, shown above or below each calibration point, are calculated using the negated absolute distance from the decision boundary as an NCM. The shaded regions capture points which are \textit{more nonconform} with respect to the predicted class (blue for class~\CIRCLE{} and red for class~\Circle{}). The alpha assessment  (d) shows the distribution of p-values and per-class thresholds derived from Q1 of the correctly classified points (see \cref{sec:search} for a discussion of more complex search strategies for finding thresholds).}
    \label{fig:toy-thresholding}
\end{figure*}

\begin{figure*}[t!]
    \centering
    \begin{subfigure}[t]{0.3\textwidth}
        \centering
        \includegraphics[height=1.3in]{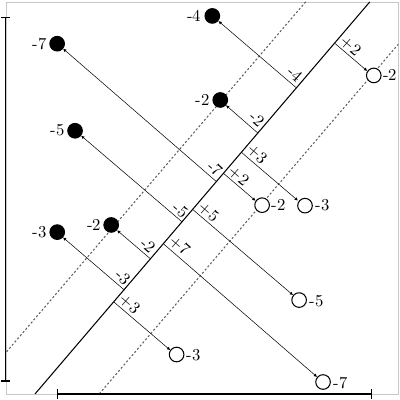}
        \caption{Distances and NCMs}
        \label{fig:test-one}
    \end{subfigure}%
    ~
    \begin{subfigure}[t]{0.3\textwidth}
        \centering
        \includegraphics[height=1.3in]{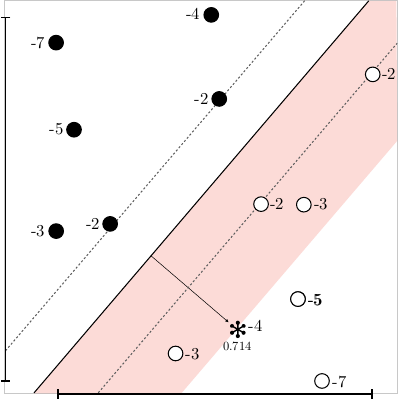}
        \caption{P-value of new point}
        \label{fig:test-two}
    \end{subfigure}%
    ~
    \begin{subfigure}[t]{0.33\textwidth}
        \centering
        \includegraphics[height=1.3in]{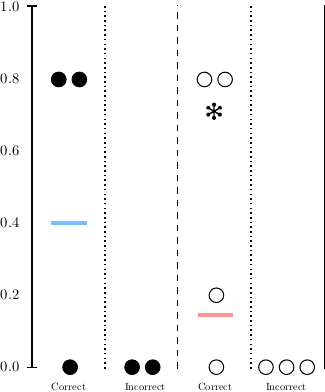}
        \caption{Comparison to threshold}
        \label{fig:test-three}
    \end{subfigure}
    \caption{Test-time procedure applied to a linear SVM and calibrated \transcend with distances from hyperplane and corresponding nonconformity scores shown in (a). In (b) a new test point is classified as class~\Circle{}. The p-value is calculated as the proportion of points belonging to \Circle{} with equal or greater nonconformity scores (captured by the shaded region) than the new point. In (c), the new point is compared against the threshold for class~\Circle{} as derived during the calibration phase (\autoref{fig:toy-thresholding}). As the p-value of the new point is greater than the threshold for the predicted class, the prediction is accepted.}
	\label{fig:toy-testing}
\end{figure*}

\section{Sound and Practical \tool}
\label{sec:transcendframework}

Once p-values are calculated, thresholds are derived to decide when to accept or reject new test examples. Here we revise and formalize the strategy used in \transcend and propose a more efficient search strategy.

\subsection{Calibration Phase}

The first phase of \transcend is the calibration procedure which searches for a set of per-class \textit{credibility} thresholds \[\mathcal{T} = \{\, \tau_y \in [0,1] : y \in \mathcal{Y} \,\}\] with which to separate drifting from non-drifting points. Given that low credibility represents a violation of conformal prediction's assumptions, these points are likely to be misclassified by the underlying classifier that similarly relies on the i.i.d. assumption. Note that thresholds can be found with different optimization criteria and it is also possible to threshold on a combination of credibility and \textit{confidence} (see~\cref{sec:credconfeval}).

Calibration aims to answer the question: ``how low a credibility is too low?'', by analyzing the p-value distribution of points in a representative, preferably stationary, environment such as the training set. Which points are selected as calibration points depends on the underlying conformal evaluator, and this comes with various trade-offs~(see \cref{sec:novel-ce}). Typically, each calibration point (or partition of the calibration set) is held out and the underlying classifier trained on the remaining points. Then a class is predicted for the calibration point(s) with p-values calculated with respect to that predicted class. This process is repeated until all calibration points are assigned a corresponding p-value. Using the ground truth, these p-values can be partitioned into \textit{correct} and \textit{incorrect} predictions that should be separated by $\mathcal{T}$. Methods to find an effective $\mathcal{T}$ can be manual (\eg picking a quartile visually using an alpha assessment) or automated (\eg grid search).

\autoref{fig:toy-thresholding} shows an example of the \transcend thresholding procedure on a toy dataset composed of two classes: \CIRCLE{} and \Circle{}. A linear SVM is paired with a TCE~(\cref{sec:tce}) to generate NCMs and p-values for the binary classification with rejection task. The decision boundary is depicted as a solid line and margins are drawn through support vectors with dotted lines. Due to the use of approximate TCE, the dataset is partitioned into folds, where each fold leaves out four points for calibration and trains on the remainder. The three folds are depicted in Figures~\ref{fig:thresh-one}, \ref{fig:thresh-two}, and \ref{fig:thresh-three}. Calibration points are shown with dotted outlines and are faded for class~\CIRCLE{}.

In each fold, a p-value is calculated for each calibration point as the proportion of other objects that are \textit{at least as dissimilar} to the predicted class as the calibration point itself. In the linear SVM setting shown, less similar objects are those closest to the decision boundary (\ie those with a higher NCM) residing in the shaded area between the decision boundary and the parallel line intersecting the point (blue for class~\CIRCLE{} and red for class~\Circle{}). The calculated p-values are shown aligned above or below each calibration point.

To evaluate how appropriate an NCM is for a given model, the p-values can be analyzed with an \textit{alpha assessment}. Here the distribution of p-values for each class are divided into groups depending on whether the calibration point was correctly or incorrectly predicted as that class. Given that there may not be enough incorrectly classified examples to perform the assessment with, it is standard to perform an alpha assessment in a \textit{non-label-conditional} manner, using p-values computed with respect to all classes, not just each point's predicted class. The greater the margin separating the distributions of correct and incorrect p-values, the better suited an NCM is for a model. The alpha assessment in \autoref{fig:thresh-alpha} shows the distribution of p-values for correctly and incorrectly predicted calibration points for classes~\CIRCLE{} and~\Circle{}. Given the size of the example dataset, the assessment is computed in a \textit{label-conditional} manner and the threshold is set at Q1 of the p-values for correctly classified points (more insight into threshold search strategies can be found in~\cref{sec:search}). Test points generating p-values below this threshold will be rejected.

\subsection{Test Phase}

At test time, there are $|\mathcal{Y}| + 1$ outcomes. When a new test object $z^*$ arrives, its p-value $p_{z^*}^{\hat{y}}$ is calculated with respect to the predicted class $\hat{y}$ (label conditional). If $p_{z^*}^{\hat{y}} < \tau_{\hat{y}}$, the threshold for the predicted class, then the null hypothesis---that $z^*$ is drifting relative to the training data and does not belong to $\hat{y}$---is approved and the prediction rejected. If $p_{z^*}^{\hat{y}}\ge \tau_{\hat{y}}$, the prediction is accepted and the object classified as $\hat{y}$.

\autoref{fig:toy-testing} follows on from the calibration example above. \autoref{fig:test-one} illustrates the NCM being used: the negated absolute distance from the hyperplane. In \autoref{fig:test-two}, a new test example \Test{} appears and is classified as class \Circle{}. The p-value $p_{\Test{}}^{\tinywc{}} = 0.714$ is calculated as the proportion of points belonging to \Circle{} with equal or greater nonconformity scores than \Test{}. Finally, \autoref{fig:test-three} shows $p_{\Test{}}^{\tinywc{}}$ compared against the threshold $\tau_{\tinywc{}}$ and, as  $p_{\Test{}}^{\tinywc{}} \ge \tau_{\tinywc{}}$, the prediction is accepted.

\subsection{Rejection Cost}

What happens to rejected points depends on the rest of the detection pipeline.
In a simple setting, rejected points may be manually inspected and labeled by specialists. Alternatively, they may continue downstream to further automated analyses or to other ML algorithms such as unsupervised systems.

In all cases there will be some cost associated with rejecting predictions. When choosing rejection thresholds, it is vital to keep this cost in mind and weigh it against the potential performance gains.
The \tesseract framework~\cite{pendlebury2019tesseract} defines three important metrics to use when tuning or evaluating a system for mitigating time decay.

\textit{Performance} ensures that robustness against concept drift is measured appropriately depending on the end goal (\eg high \fone score or high TPR at an acceptable FPR threshold).

\textit{Quarantine cost} measures the cost incurred by rejections. This is important for putting the performance of kept elements in perspective---there will often be a trade-off between the amount of rejections and higher performance on kept points.

\textit{Labeling cost} measures the manual effort needed to find ground truth labels for new points. While this is more pertinent to retraining strategies, it is related to the overhead associated with rejection as many may need to be manually labeled. As an example,~\citet{miller16dimva} estimate that the labeling capacity for an average company is 80 samples per day.

\subsection{Improving the Threshold Search}
\label{sec:formalize}
\label{sec:search}

Here we model the calibration procedure as an optimization problem for maximizing a given performance metric (\eg \fone, Precision, or Recall of kept elements). Usually this maximization is subject to some constraint on another metric, for example, it is trivial to attain high \fone performance in kept elements by accepting very few high quality predictions, but this \revised{will cause many correct predictions to be rejected.}

Formally, given $n$ calibration points, we represent this as:

\begin{equation}
\begin{aligned}
\argmax_{\mathcal{T}} \quad & \mathcal{F}(\bm{Y}, \bm{\hat{Y}}, P; \mathcal{T})\\
\textrm{subject to} \quad & \mathcal{G}(\bm{Y}, \bm{\hat{Y}}, P; \mathcal{T}) \ge \mathcal{C} \,,\\
\end{aligned}
\end{equation}
where \(\bm{Y}\) and \(\bm{\hat{Y}}\) are $n$-dimensional vectors of ground truth and predicted labels respectively, $P$ is a \(|\mathcal{Y}| \times n\)-dimensional matrix of calibration p-values, and \(\mathcal{T} = \{\, \tau_y \in [0,1] \mid y \in \mathcal{Y} \,\}\) is the set of thresholds.
The objective function \(\mathcal{F}\) maps these inputs to the metric of interest in \(\mathbb{R}\), for example \fone of kept elements, while \(\mathcal{G}\) maps to the metric to be constrained, such as the number of per-class rejected elements. \(\mathcal{C}\) is the threshold value that bounds the constraint function.

\begin{figure*}[t!]
    \centering
    \begin{subfigure}[t]{0.13\textwidth}
        \centering
        \includegraphics[height=0.5in]{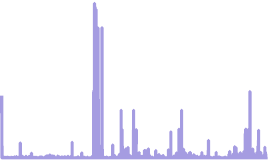}
        \caption{Training \citep{jordaney2017transcend}}
        \label{fig:usenix-dist-0}
    \end{subfigure}%
    \hfill
    \begin{subfigure}[t]{0.13\textwidth}
        \centering
        \includegraphics[height=0.5in]{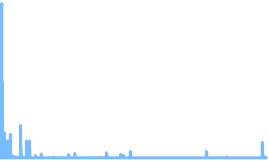}
        \caption{Test \citep{jordaney2017transcend}}
        \label{fig:usenix-dist-1}
    \end{subfigure}%
    \rulesep
    \begin{subfigure}[t]{0.13\textwidth}
        \centering
        \includegraphics[height=0.5in]{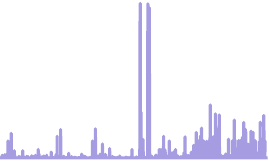}
        \caption{Training}
        \label{fig:transcend-dist-0}
    \end{subfigure}%
    \hfill
    \begin{subfigure}[t]{0.13\textwidth}
        \centering
        \includegraphics[height=0.5in]{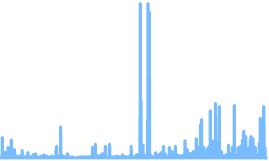}
        \caption{Test at 1 year}
        \label{fig:transcend-dist-12}
    \end{subfigure}%
    \hfill
    \begin{subfigure}[t]{0.13\textwidth}
        \centering
        \includegraphics[height=0.5in]{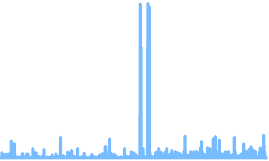}
        \caption{Test at 2 years}
        \label{fig:transcend-dist-24}
    \end{subfigure}%
    \hfill
    \begin{subfigure}[t]{0.13\textwidth}
        \centering
        \includegraphics[height=0.5in]{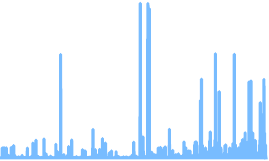}
        \caption{Test at 3 years}
        \label{fig:transcend-dist-36}
    \end{subfigure}%
    \hfill
    \begin{subfigure}[t]{0.13\textwidth}
        \centering
        \includegraphics[height=0.5in]{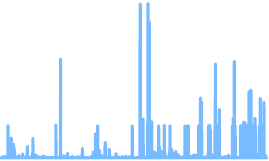}
        \caption{Test at 4 years}
        \label{fig:transcend-dist-48}
    \end{subfigure}
    \caption{Frequency distributions of features depicting covariate shift between training and test malware examples. The data from~\citet{jordaney2017transcend}, displayed in (a) and (b), shows a sudden and significant shift, while the data used in~\cref{sec:eval}, displayed in (c--g), shows a more subtle, natural drift occurring over time.}
    \label{fig:distributions}
\end{figure*}

Given this formalization, we propose an alternative random search strategy to replace the exhaustive grid search used in the original paper~\cite{jordaney2017transcend}. In the exhaustive grid search, each possible combination of thresholds over all classes is tested systematically, considering some fixed range of variables \(V = \{v : v \in [0,1]\}\). However, this suffers from the curse of dimensionality~\cite{bellman61curse}, resulting in \(|V|^{|\mathcal{Y}|}\) total trials, growing exponentially with the number of classes. Additionally, reducing the granularity of the range considered in \(V\) increases the risk of `skipping' over an optimal threshold combination. Similarly, often many useless threshold combinations are considered (where one is either too high or too low). This failure to evenly cover subspaces of interest worsens as the dimensionality increases~\cite{bergstra12search}, making it especially problematic for multiclass classification. The granularity for \(V\) can be chosen manually based on intuition, however this results in parameters which are difficult to reproduce and transfer to other settings.

It has been shown for hyperparameter optimization that random search is able to find combinations of variables at least as optimal as those found with full grid search over the same domain, at a fraction of the computational cost~\cite{bergstra12search}. We apply these findings to the threshold calibration and replace the exhaustive grid search with a random search (\Cref{alg:search}). We choose random combinations of thresholds in the interval \([0,1]\), keeping track of the thresholds that maximize our chosen metric given the constraints (see~\Cref{sec:formalize}). The search continues until either of two conditions are met. A limit is set on the number of iterations, determined by the time and resources that are available for the calibration. Intuitively a higher limit will increase the likelihood of finding better thresholds and so acts as the upper bound of the optimization. Secondly, a stop condition can be set. In this work we consider a \textit{no-update} approach in which the search will stop once a fixed point is found, \ie if there is no improvement to performance after a certain number of consecutive iterations. Note that this search procedure can be easily parallelized.

We empirically compare the two search strategies in~\Cref{sec:search-eval}.

\section{Experimental Evaluation}
\label{sec:eval}

We evaluate our novel evaluators when faced with gradual concept drift caused by the evolution of real malicious Android apps over time (\cref{sec:conf-eval}), the performance gained by including confidence scores (\cref{sec:credconfeval}), how our random search implementation fares against exhaustive search (\cref{sec:search-eval}),
\revised{how the evaluators compare to alternative methods~(\cref{sec:prior-methods}), and perform on PE and PDF malware domains~(\cref{sec:beyond-android}).
}

\subsection{Experimental Settings}

\paragraph{Prototype} We implement \tool as a Python library encompassing out new proposals as well as the functionality of the original \transcend. %
 We release the code as open source---note that this is the first publicly available implementation of \transcendnocite in any form.

\paragraph{Dataset} We first focus on malware detection in the Android domain. We sample 232,848 benign and 26,387 malicious apps
from \az~\cite{androzoo}. This allows us to demonstrate efficacy when faced with a natural, surreptitious concept drift. The apps span 5 years, from Jan 2014 through to Dec 2018.
We use the \tesseract~\cite{pendlebury2019tesseract} framework to temporally split the dataset, ensuring that \tesseract's constraints are accounted for to remove sources of spatial and temporal experimental bias. Training and calibration are performed using apps from 2014 and testing is evaluated over the concept drift that occurs over the remaining period on a month-by-month basis.

\paragraph{Eliminating Sampling Bias} The original evaluation of \transcend artificially simulated concept drift by fusing two datasets: \text{Drebin}~\cite{arp2014drebin} and \text{Marvin}~\cite{marvin}, a process which may have induced experimental bias~\cite{arp2020dodo} and made it easier to detect drifting examples.  \autoref{fig:distributions} shows a visibly significant covariate shift in the distribution of features for training and test malware examples from~\citet{jordaney2017transcend}, with a Kullback-Leibler (KL) divergence~\cite{kullback1951}---an unbounded measure of distribution difference---of 696.66. The covariate shift in our dataset is much more subtle and natural over time, with an average KL divergence of 189.55 between each training and test partition. From this we conclude that the distributions were significantly more different in the original evaluation than would be expected in naturally occurring concept drift, which would have made it easier to detect drifting examples.

\paragraph{Classifier}
For the underlying classifier, we use \text{Drebin}~\cite{arp2014drebin} which has been shown to achieve state-of-the-art performance if a retraining strategy is used to remediate concept drift~\cite{pendlebury2019tesseract}. Due to this, we hypothesize that if \transcend is used to reject drifting points, \text{Drebin} will be able to classify the remaining points with high accuracy. \text{Drebin} uses a linear SVM and a binary feature space where components (activities, permissions, URLs, etc) are represented as present or absent.

\paragraph{Calibration} To optimize the thresholding, we maximize the \fone of all kept elements for a rejection rate less than 15\%. These metrics are computed in aggregate for each time period of the temporal evaluation. On our dataset, this would amount to an average rejection of $\sim$20 samples a day, well below the estimated labeling capacity of 80 a day suggested by~\citet{miller16dimva}. \revised{However, we note that these constraints may need to be adjusted according to specific operational requirements, for example, it may be more appropriate to minimize the rejection rate while sacrificing \fone for kept elements}. For the random search we use 100,000 random iterations with early stopping after 3,000 consecutive events without improvement. For approx-TCE and CCE we calibrate using \(k=10\) folds.

\subsection{Novel Conformal Evaluators}
\label{sec:conf-eval}

\begin{figure*}[t!]
    \centering

        \centering
        \includegraphics[width=\textwidth]{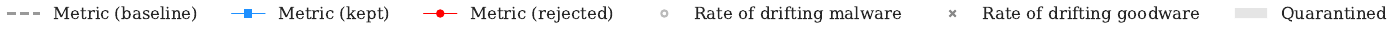}
        \label{fig:legend}
    \begin{subfigure}[t]{0.3\textwidth}
        \centering
        \includegraphics[width=\textwidth]{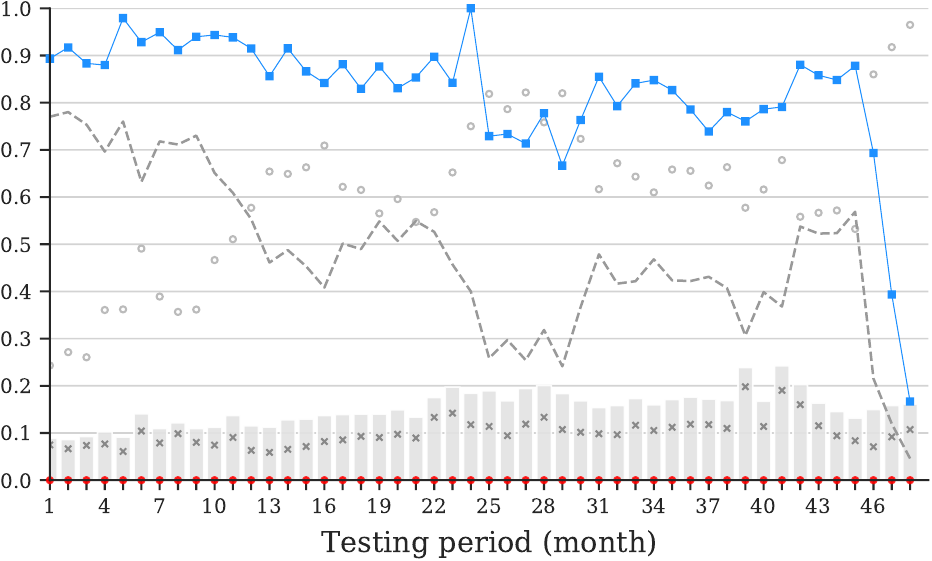}
        \caption{$F_1$-Score, Approx-TCE w/ credibility}
        \label{fig:fpr-f1-tce}
    \end{subfigure}
    \hfill
    \begin{subfigure}[t]{0.3\textwidth}
        \centering
        \includegraphics[width=\textwidth]{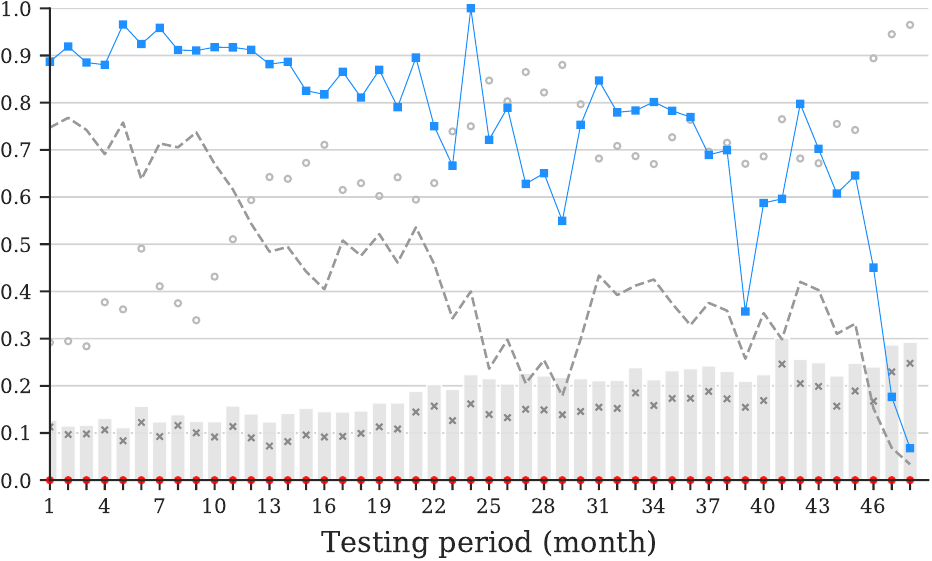}
        \caption{$F_1$-Score, ICE w/ credibility}
        \label{fig:fpr-f1-ice}
    \end{subfigure}%
    \hfill
    \begin{subfigure}[t]{0.3\textwidth}
        \centering
        \includegraphics[width=\textwidth]{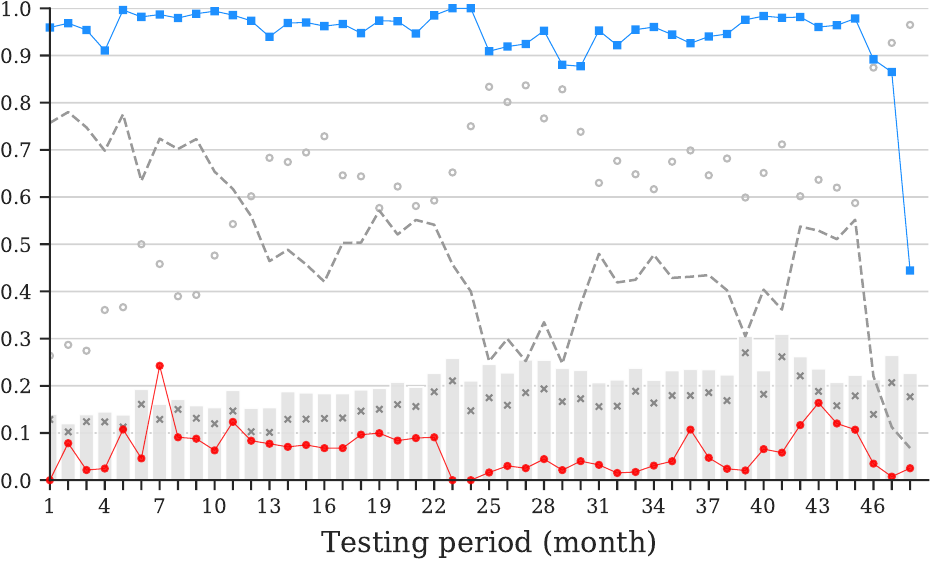}
        \caption{$F_1$-Score, CCE w/ credibility}
        \label{fig:fpr-f1-cce}

    \vspace{0.8em}
    \end{subfigure}

    \begin{subfigure}[t]{0.3\textwidth}
        \centering
        \includegraphics[width=\textwidth]{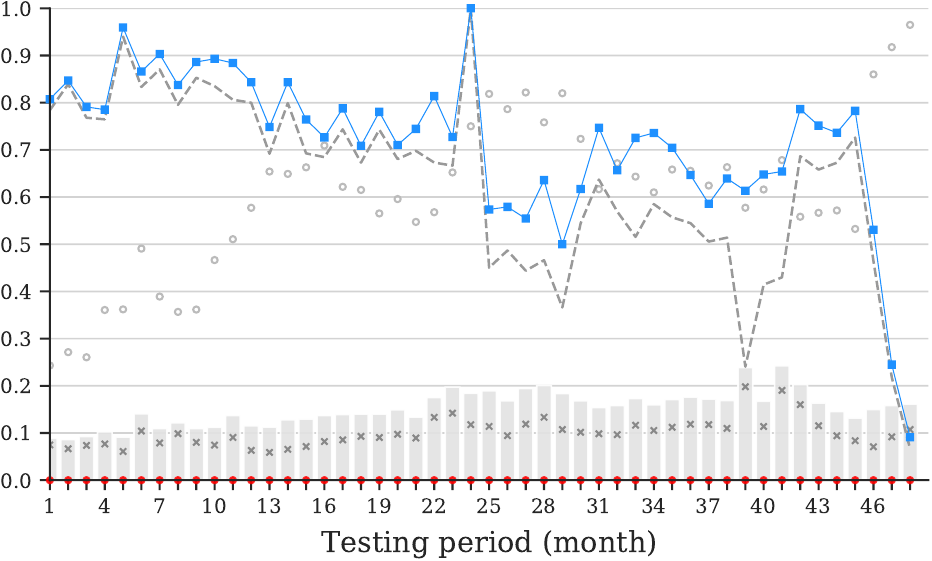}
        \caption{Precision, Approx-TCE w/ credibility}
        \label{fig:fpr-precision-tce}
    \end{subfigure}%
    \hfill
    \begin{subfigure}[t]{0.3\textwidth}
        \centering
        \includegraphics[width=\textwidth]{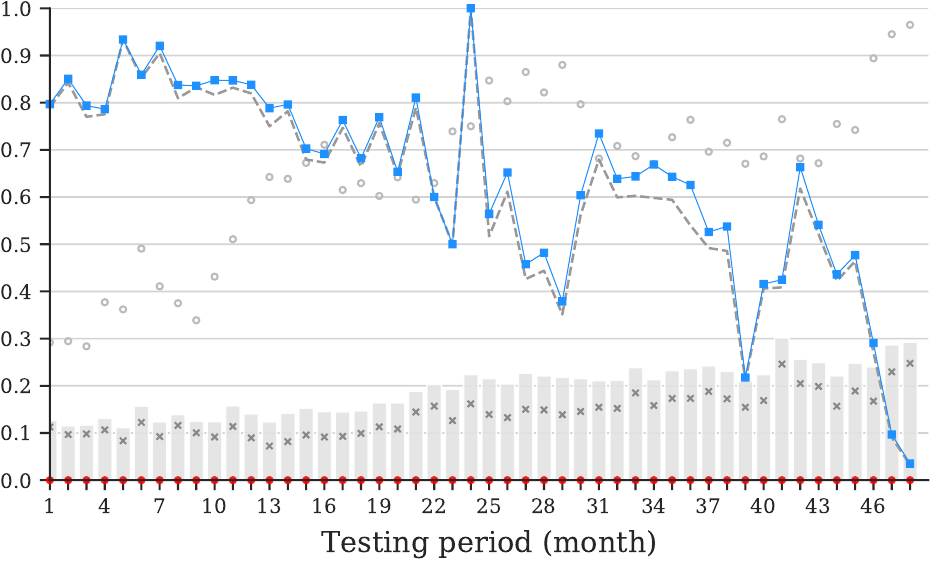}
        \caption{Precision, ICE w/ credibility}
        \label{fig:fpr-precision-ice, credibility}
    \end{subfigure}%
    \hfill
    \begin{subfigure}[t]{0.3\textwidth}
        \centering
        \includegraphics[width=\textwidth]{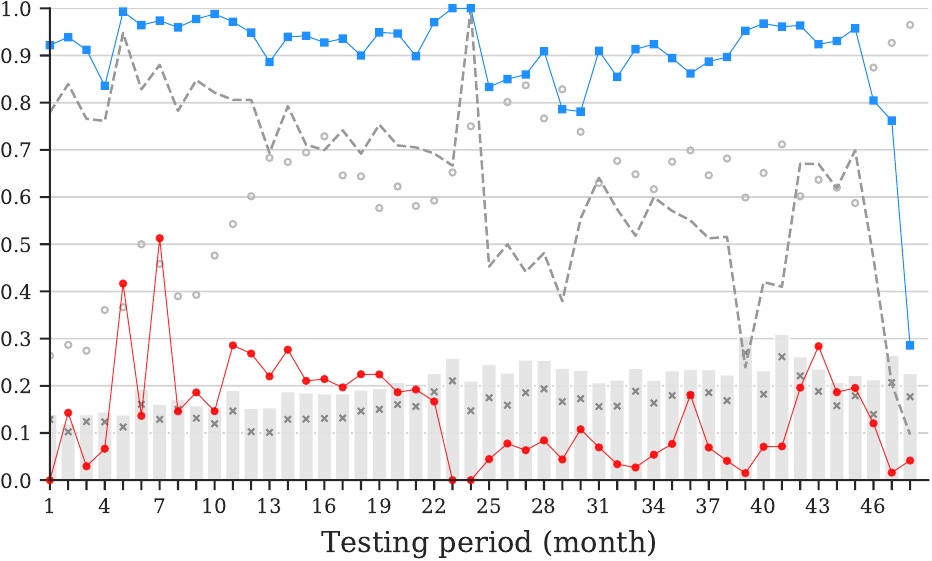}
        \caption{Precision, CCE w/ credibility}
        \label{fig:fpr-precision-cce, credibility}

    \vspace{0.8em}
    \end{subfigure}

    \begin{subfigure}[t]{0.3\textwidth}
        \centering
        \includegraphics[width=\textwidth]{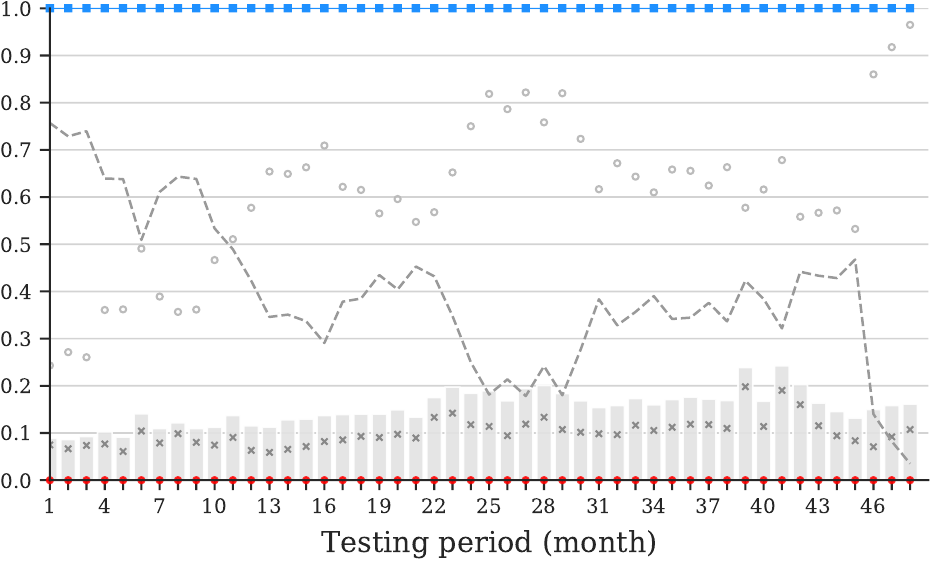}
        \caption{Recall, Approx-TCE w/ credibility}
        \label{fig:fpr-recall-tce}
    \end{subfigure}%
    \hfill
    \begin{subfigure}[t]{0.3\textwidth}
        \centering
        \includegraphics[width=\textwidth]{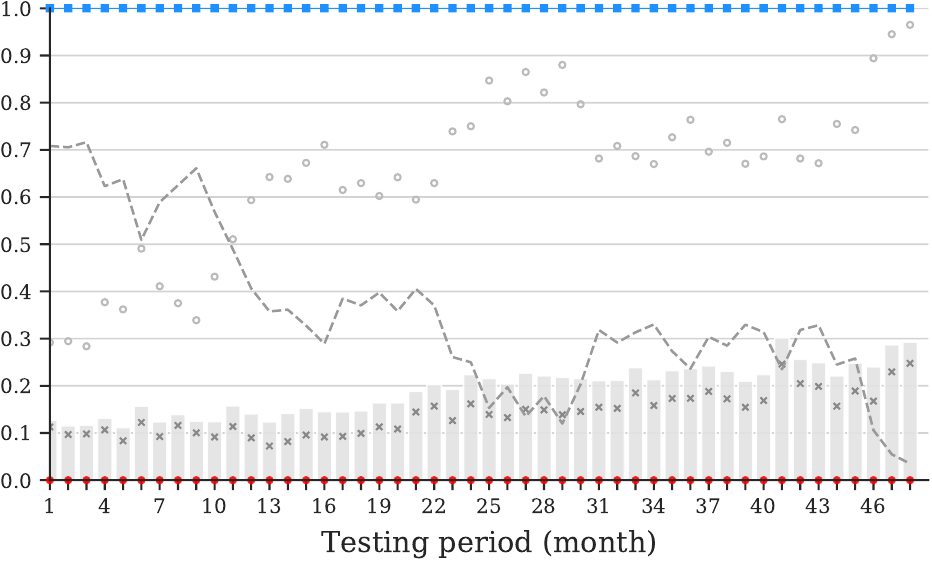}
        \caption{Recall, ICE w/ credibility}
        \label{fig:fpr-recall-ice}
    \end{subfigure}%
    \hfill
    \begin{subfigure}[t]{0.3\textwidth}
        \centering
        \includegraphics[width=\textwidth]{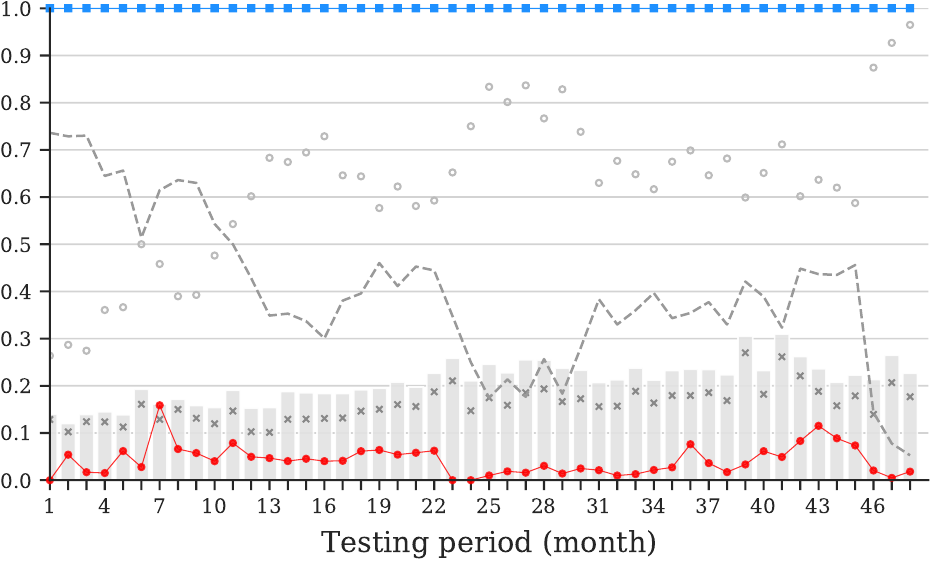}
        \caption{Recall, CCE w/ credibility}
        \label{fig:fpr-recall-cce}

    \vspace{0.3em}
    \end{subfigure}

        \centering

	\color{lightgray}{
    \begin{center}
        \begin{tikzpicture}
            \draw[*-*, line width=0.1mm] (0,0) to (0.99\linewidth,0);
        \end{tikzpicture}
    \end{center}
    }
	\vspace{0.7em}

    \begin{subfigure}[t]{0.3\textwidth}
        \centering
        \includegraphics[width=\textwidth]{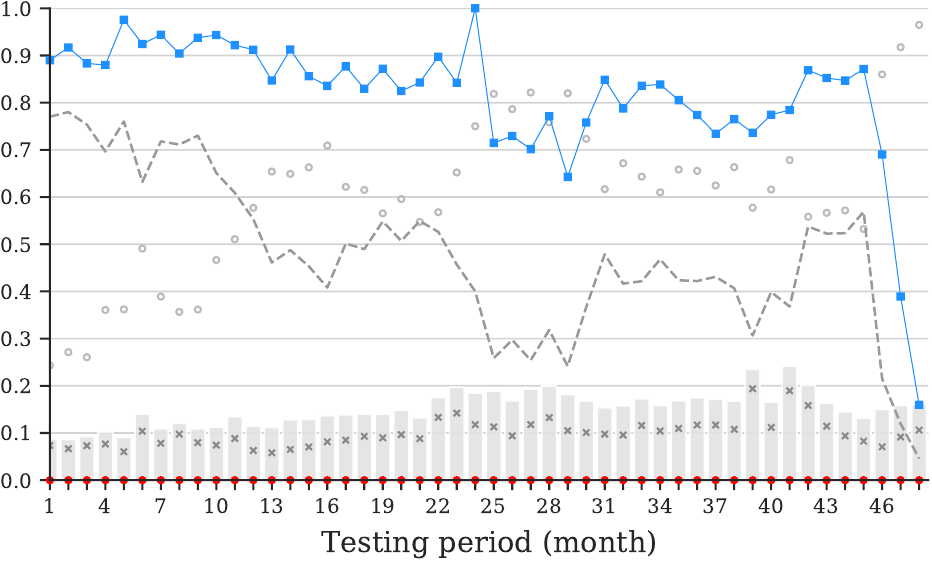}
        \caption{\fone-Score, Approx TCE w/ cred + conf}
        \label{fig:f1-units-tce-credconf}
    \end{subfigure}%
    \hfill
    \begin{subfigure}[t]{0.3\textwidth}
        \centering
        \includegraphics[width=\textwidth]{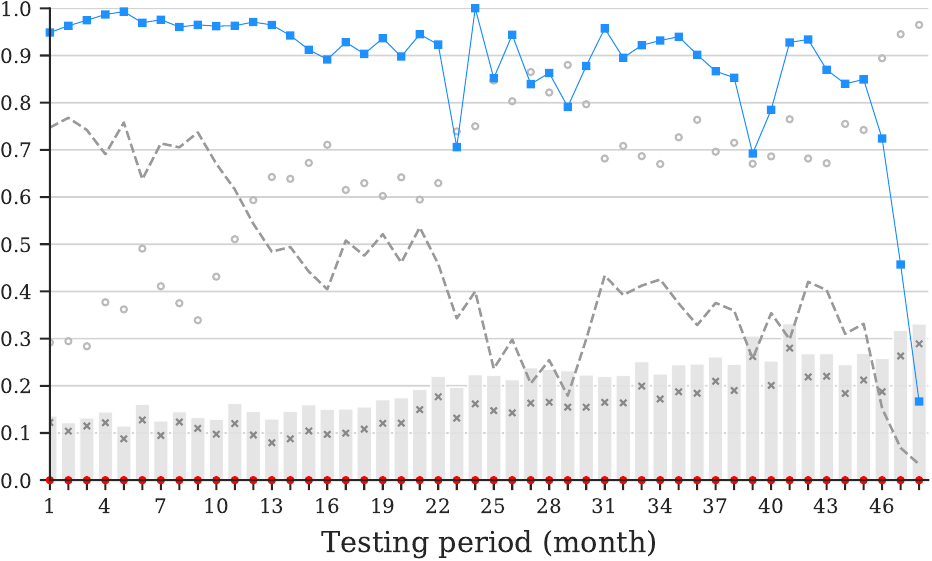}
        \caption{\fone-Score, ICE w/ cred + conf}
        \label{fig:f1-units-ice-credconf}
    \end{subfigure}%
    \hfill
    \begin{subfigure}[t]{0.3\textwidth}
        \centering
        \includegraphics[width=\textwidth]{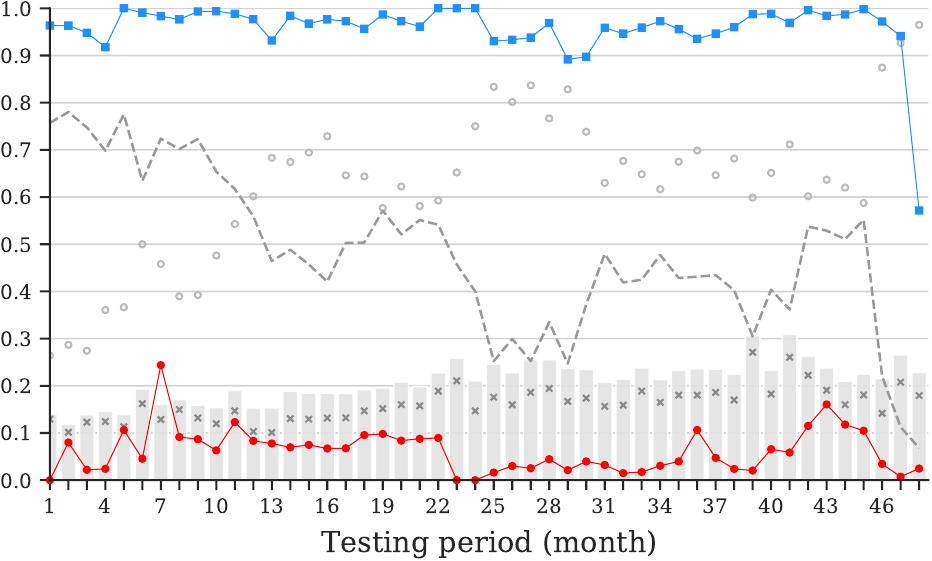}
        \caption{\fone-Score, CCE w/ cred + conf}
        \label{fig:f1-units-cce-credconf}
    \vspace{0.8em}
    \end{subfigure}

    \begin{subfigure}[t]{0.3\textwidth}
        \centering
        \includegraphics[width=\textwidth]{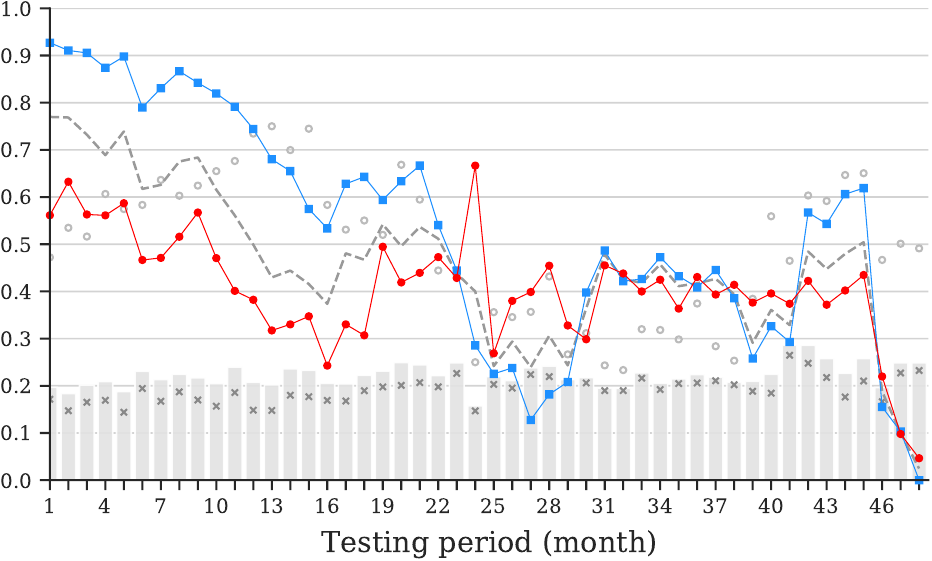}
        \caption{\fone-Score, Approx-TCE w/ probabilities}
        \label{fig:f1-units-tce-prob}
    \end{subfigure}%
    \hfill
    \begin{subfigure}[t]{0.3\textwidth}
        \centering
        \includegraphics[width=\textwidth]{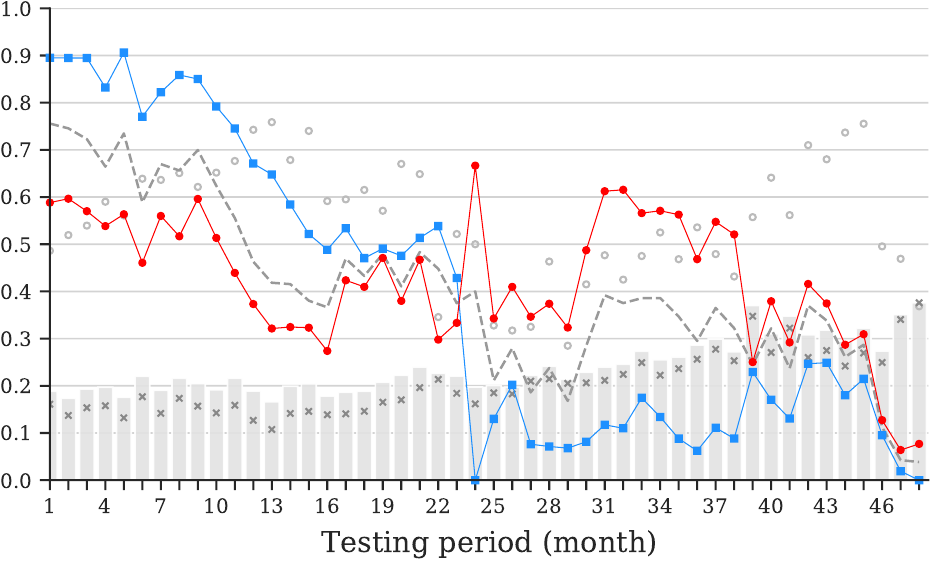}
        \caption{\fone-Score, ICE w/ probabilities}
        \label{fig:f1-units-ice-prob}
    \end{subfigure}%
    \hfill
    \begin{subfigure}[t]{0.3\textwidth}
        \centering
        \includegraphics[width=\textwidth]{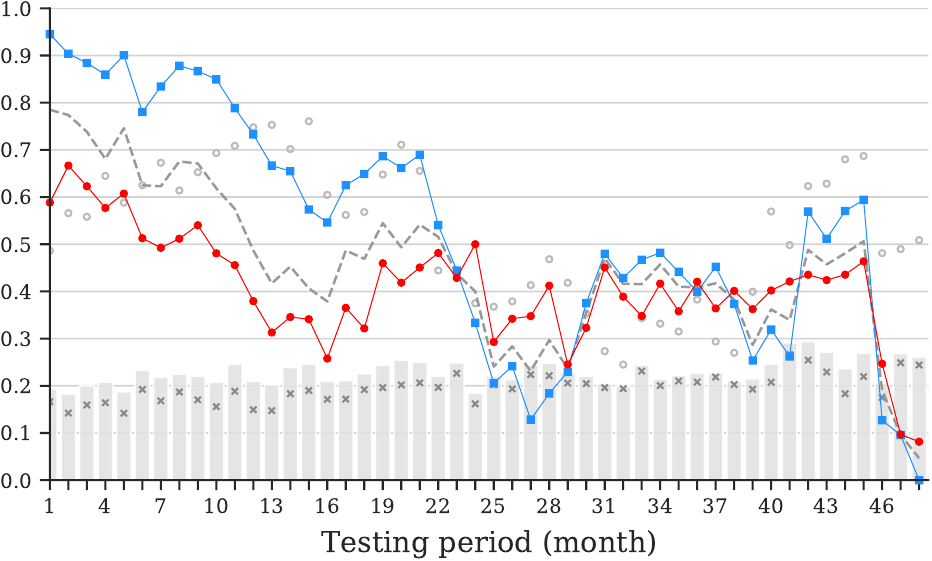}
        \caption{\fone-Score, CCE w/ probabilities}
        \label{fig:f1-units-cce-prob}

    \vspace{0.5em}
    \end{subfigure}

    \caption{Performance for the three proposed conformal evaluators (columns) using different quality metrics. The first three rows show \fone-Score, Precision, and Recall, respectively, of the different evaluators using credibility p-values. The lower two rows show \fone-Score using a combination of credibility and confidence p-vaules, and probabilities, respectively. The dashed line shows the performance with no rejection mechanism. The upper line (\(\square\) marker) shows the performance on kept examples whose classifications were accepted. The lower line (\Circle marker) shows the performance on rejected examples. These are the mistakes that would have been made if the predictions were accepted by the degrading model. The bars show the proportion of rejected elements in each period, while the x and o markers show the proportion of \textit{ground truth} malware and goodware that was identified as drifting and quarantined, respectively.}
    \label{fig:fpr-comparison}
\end{figure*}
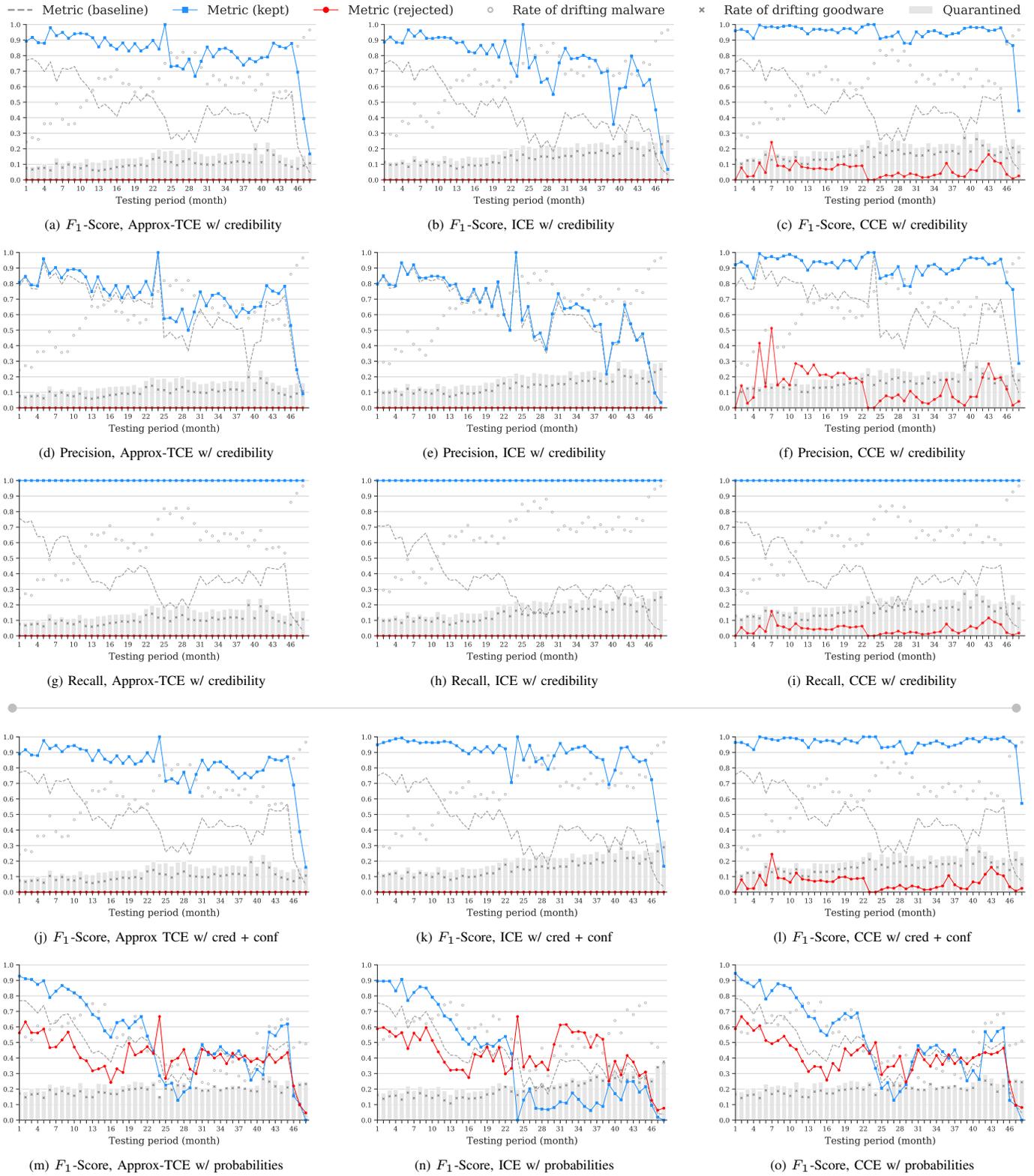

Here we compare the novel conformal evaluators of \tool. As vanilla TCE is not feasible for this experiment setting due to the size of the training set~(\cref{sec:tce}), we use approx-TCE as a stand-in, \revised{however we provide a minimal experiment in~\Cref{app:full-tce} to show that the expected performance difference between vanilla TCE and our evaluators is negligible.}

\paragraph{Performance Metrics} \Cref{fig:fpr-comparison} shows the the \fone, Precision, and Recall (rows 1--3) for each of the novel evaluators (columns). The middle dashed line shows the baseline performance when no rejection is enforced. This is the performance decay caused by concept drift present in the dataset that results from an evolving malicious class. Note that classifiers degrade rapidly, becoming worse than random in under one year.

The upper solid line shows the performance of kept elements, those with test p-values that fall above the threshold of their predicted classes. While decay is still present, approx-TCE and ICE are able to maintain \fone \(>\) 0.8 for two years, doubling the lifespan of the model.
Note that the sudden drop in performance of the last three months is likely an artifact of the fewer examples crawled by \az in those months.

The lower solid line shows the performance of rejected elements. Low metrics mean the rejected elements would have been incorrectly classified by the underlying classifier and were rightfully rejected, while high metrics means rejections were erroneous. Approx-TCE and ICE have \fone, Precision, and Recall of 0 for rejected elements for every test month meaning that all rejected elements would have been misclassified.

The result of CCE differs in that it is less conservative in its rejections. The performance of kept elements is much higher, but also slightly higher for rejected elements, indicating that a small proportion of rejected elements would have actually been correctly classified. We observe that this conservatism can be increased or decreased by modifying the conditions of the majority vote. If more folds are required to agree before a decision is accepted, the CCE will be more conservative, rejecting more elements. If less folds are required, more elements will be accepted. In this respect, CCE offers an alternative dimension of tuning in addition to the threshold optimization. Additionally, this is parameter can be altered during a deployment, rather than being set at calibration. This allows for some adaptability, such as when the cost of False Negatives is very high (\eg not alerting security teams to attacks in network intrusion detection), or when the cost of False Positives is very high (\eg withholding benign emails in spam detection, or disabling legitimate user accounts in fake account detection). A further empirical analysis of the effect of the majority vote conditions is included in~\Cref{app:cce-tuning}.

\paragraph{Rejection Rates} Gray bars show the proportion of rejected test elements.
In each case, rejections begin close to the rate used for calibration before slowly rising each year, averaging 26.45 samples per day.
While rejection rates may appear high, these are symptomatic of rising concept drift and deteriorating performance in the underlying classifier and are often preferable to taking incorrect actions on False Positives and False Negatives. In an extreme case where a classifier always predicts the incorrect label, rejection rates could reach 100\% but the \fone of rejected elements would be 0\%.
\revised{The gray markers show the proportion of ground truth malware and goodware that are rejected each period, illustrating the evaluators' perception of drift in that class. Strikingly, for our evaluators the drift rate of the malicious class is \textit{inversely correlated} to the performance loss in the baseline, while the drift rate for goodware is relatively stable. This supports our hypothesis that performance decay is mostly driven by evolution in the malicious class. We reiterate that in the case of Approx-TCE and ICE, the low \fone of rejected elements indicates that \textit{all} of the rejected malware would have evaded the classifier if they had not been identified as drifting.}

\paragraph{Runtime} The runtime of the conformal evaluators during this experiment match what would be expected from their relative complexities (\cf~\Cref{tab:complexities}). The ICE is the quickest at 11.5 CPU hours. The CCE took 35.6 CPU hours, but our implementation is parallelized resulting in a wall-clock time identical to the ICE. The Approx-TCE took 46.1 CPU hours. As discussed, vanilla TCE was computationally infeasible, but we estimate a runtime of 1.9 CPU years, considering that the time required to fit the underlying classifier once is $\sim$10 minutes \revised{and the classifier must be trained once \textit{for each} training example.}

We conclude that the ICE is the most useful for settings where resources are limited or models with a rapid iteration cycle (\eg daily), while the CCE offers greater confidence and flexibility at a slightly higher computational cost.

\begin{table*}[t]
\caption{Area Under Time (AUT) of \fone performance with respect to concept drift over the 48 month test period for different quality metrics: credibility, credibility with confidence, and probabilities (\cf~\Cref{fig:fpr-comparison}). We aim to \textit{maximize} the metrics of kept elements and \textit{minimize} the metrics for rejected elements.}
    \footnotesize
    \centering
    \begin{tabular}{llP{1.65cm}P{1.65cm}P{1.65cm}}
    \toprule
                        &   & \textbf{Approx-TCE} & \textbf{ICE} & \textbf{CCE} \\
    \midrule
    \multirow{3}{*}{Baseline}
    & AUT(\fone w/ credibility, 48\textit{m})     & .480  & .440  & .483           \\
    & AUT(\fone w/ cred + conf, 48\textit{m})     & .480  & .440  & .483           \\
    & AUT(\fone w/ probability, 48\textit{m})     & .456  & .405  & .455           \\
    \midrule
    \multirow{3}{*}{Kept Elements}
    & AUT(\fone w/ credibility, 48\textit{m})     & .829  & .762  & .950           \\
    & AUT(\fone w/ cred + conf, 48\textit{m})     & .822  & .887  & .962           \\
    & AUT(\fone w/ probability, 48\textit{m})     & .531  & .388  & .532           \\
    \midrule
    \multirow{3}{*}{Rejected Elements}
    & AUT(\fone w/ credibility, 48\textit{m})     & .000  & .000  & .064           \\
    & AUT(\fone w/ cred + conf, 48\textit{m})     & .000  & .000  & .063           \\
    & AUT(\fone w/ probability, 48\textit{m})     & .410  & .426  & .410           \\
    \bottomrule
\end{tabular}

\label{tab:aut-eval}
\end{table*}

\subsection{Credibility, Confidence, and Probabilities}
\label{sec:credconfeval}

Here we compare the performance under different quality metrics. The exact performance over time for all settings discussed in this subsection is reported in \autoref{tab:aut-eval}.%

\paragraph{Credibility with Confidence} Intuitively, including confidence thresholds when evaluating a classifier prediction would be beneficial as confidence represents how certain the classifier is in its own prediction. However, as credibility is the main indicator that i.i.d. has been violated, and thus that concept drift is occurring, it is unclear what further gain confidence could provide. Here we test this by evaluating the conformal evaluators under the same conditions as~\Cref{sec:conf-eval}, using per-class thresholds for both credibility and confidence.

\Cref{fig:fpr-comparison} compares the \fone for each conformal evaluator (columns) using different thresholding metrics (rows~1, 4--5). The upper blue line shows the performance of kept elements while the lower red line shows the performance of rejected elements. The gray dashed line depicts the baseline performance when no rejection mechanism is used.

The penultimate row shows the \fone when confidence is included. Performance for the approx-TCE and CCE is relatively unchanged, but is markedly improved for the ICE with degradation postponed much longer than before. The confidence appears to restore some of the statistical information lost by using only a small amount of the training data for calibration.

However, the computation required to find thresholds is higher than with credibility only---equivalent to doubling the number of classes. We conclude that the performance gain from including confidence is situationally dependent; although it will improve the accuracy of an ICE, a CCE will often provide the same accuracy with comparable calibration time.

\paragraph{Probabilities} The final row of \Cref{fig:fpr-comparison} shows the \fone when the classifiers' output probabilities are used for thresholding, rather than generating per-class p-values for each calibration and test point. For each evaluator, the same training and calibration split is used as with p-values, to ensure a fair comparison. The plot shows probabilities alone offer a very small improvement for kept elements over the baseline in the first year, becoming increasingly volatile as the concept drift becomes more severe. \revised{Additionally, the perceived drift rate for each class has no relation to the baseline performance loss, indicating that the root cause of the drift is not identified.} This shows the statistical support offered by the conformal evaluator's p-value computation is significant and justifies the additional computational overhead that it induces.

\subsection{Full Grid Search vs Random Search}
\label{sec:search-eval}

Here we evaluate our random search implementation~(\Cref{sec:search}) compared to the full grid search used in the original proposal~\cite{jordaney2017transcend}. We show the random search can find high quality calibration thresholds more efficiently than the full search.

Due to the full grid search cost, here we train and calibrate on 1 month of data and test on the remaining 59 months using an approx-TCE with 10 folds. We maximize \fone for an acceptable rejection rate of less than 15\%. To ensure the baseline discovers high quality thresholds we use a fine granularity grid covering 1,317,520 combinations of thresholds. For random search we set an upper limit of 10,000 trials.

\Cref{tab:search} compares the performance without rejection, with rejection thresholds from the full grid search, and with rejection thresholds from random search. Note there is no significant performance difference between the two strategies, but the random search is able to cover the same search space with two orders of magnitude fewer trials. We observe that the full grid search makes erroneous assumptions on the distribution of quality thresholds which the random search does not. Additionally, while the random search allows for a variety of stopping conditions, the only mechanism to control the length of the full grid search is the size of the interval to search and the granularity of the search steps---which are difficult to choose beforehand.

\revised{
\subsection{Comparison to Prior Approaches}
\label{sec:prior-methods}
}

\revised{
To provide further context on the performance of \tool, we compare against two closely related state-of-the-art approaches: \citet{linusson2018pakdd} (which we denote \cpreject) and \droidevolver~\cite{xu2019edvolver}.
}

\revised{
\paragraph{\cpreject~\citep{linusson2018pakdd}} The first approach is a recent method for performing rejection using conformal prediction. For a given test set and hyperparameter~\(k\), \cpreject aims to output the largest possible set of predictions containing on average no more than~\(k\) errors, while rejecting test objects for which it is too uncertain. The training and calibration dataset splits are the same as we use for the ICE setting; however while \tool makes decisions on individual test objects as they appear,
\cpreject operates \textit{a posteriori} on a batch of test inputs and predictions. Given this advantage, to ensure a fair comparison we test on each month with~\(k\) set to the 85th percentile which ensures a rejection rate of 15\%---the same rejection rate \tool is calibrated for. The underlying classifier is a random forest classifier with 100 trees and the conformal prediction NCM is the maximum margin between the output probability for the predicted class and the output probabilities for all other classes.
\paragraph{\droidevolver~\cite{xu2019edvolver}} The second approach is a state-of-the-art Android malware detector designed for drift \textit{adaptation}, but that includes a rejection component, in which the drift identification mechanism is inspired by the original \transcend. \droidevolver is built on an ensemble of five linear online learners, with a weighted sum as the ensemble decision function.
 For each new test object a \textit{juvenilization indicator}~(JI) score is computed per model as the proportion of apps in a fixed-size buffer of previously encountered apps, of the same class, that have decision scores greater than the new object. An object is marked as drifting  when the JI score falls outside of a precalibrated range and the corresponding decisions are rejected, \ie excluded from the weighted sum which is used to pseudo-label and update with the drifting point. The ongoing performance of the system relies on the quality of the pseudo-labels and thus indirectly on the quality of the drift identification. The JI scores are very similar to the credibility p-values from conformal evaluation, with the computational complexity of full TCE being addressed by using the small fixed-size app buffer: drift identification should be effective so long as the app buffer is representative of the overall data population. Due to this relationship, it is informative to compare against \tool.

\begin{figure}[!tb]
    \centering
        \begin{subfigure}[t]{0.48\columnwidth}
        \centering
        \includegraphics[height=1.5in]{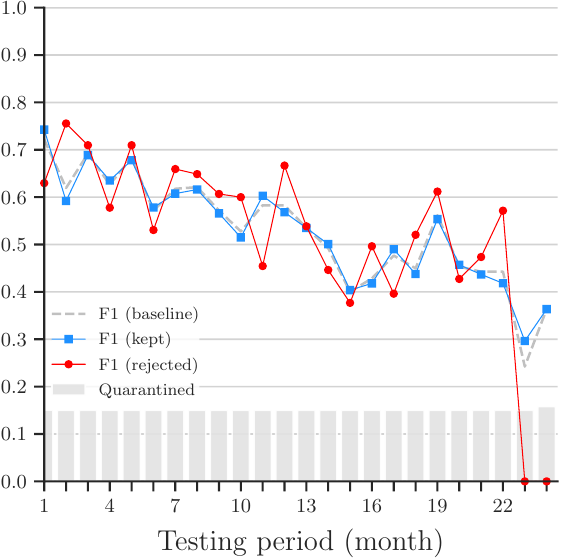}
        \caption{\cpreject\cite{linusson2018pakdd}}
        \label{fig:prior-pakdd}
    \end{subfigure}
    \hfill
    \begin{subfigure}[t]{0.48\columnwidth}
        \centering
        \includegraphics[height=1.5in]{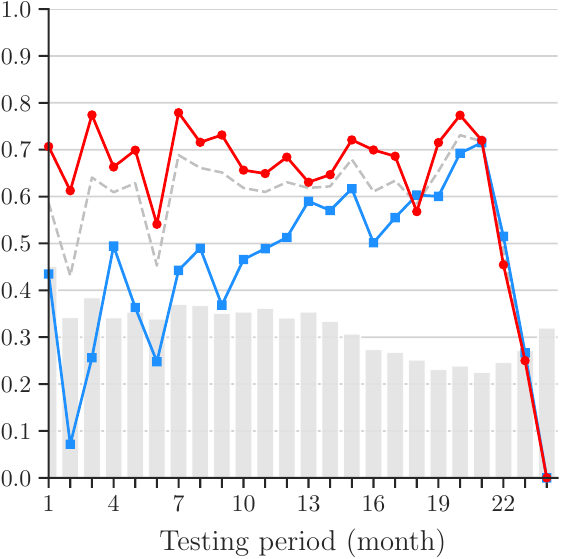}
        \caption{\droidevolver~\cite{xu2019edvolver}}
        \label{fig:prior-droidevolver}
    \end{subfigure}

    \caption{\fone-Score over time for two prior approaches with mechanisms similar to \tool (\cf \Cref{fig:fpr-f1-ice,fig:fpr-f1-cce}).}
    \label{fig:priorapproaches}
\end{figure}

\paragraph{Results} \autoref{fig:priorapproaches} shows the \fone performance of \cpreject and \droidevolver trained and calibrated on the first year of the dataset and tested on the two subsequent years at monthly intervals. This can both be compared to the first 24 months of ICE and CCE results of \Cref{fig:fpr-f1-ice,fig:fpr-f1-cce}.
For \cpreject, the similar \fone performance for kept, rejected, and baseline predictions indicate that it is unable to distinguish between drifting and non-drifting points. Although it may effectively reject low-quality predictions in a stationary environment, conformal prediction relies heavily on the exchangeability assumption, which is violated in this dataset. To obtain a prediction, \cpreject follows conformal prediction principles and outputs the class with the highest credibility~(see \Cref{sss:credconf}), but we argue this output is not trustworthy under drifting conditions. Conversely in conformal \textit{evaluation}, by decoupling the prediction of the underlying classifier from the rejection mechanism and directly interpreting the credibility as a measurement of drift when comparing it to the calibrated thresholds, we can more effectively detect poor quality predictions.
}

\revised{
While the detection performance of \droidevolver is mediocre on this dataset, the pseudo-labeling update mechanism manages to stabilize the system against the impact of drift up until the last four months. After this, performance deteriorates due to the poor quality of pseudo-labels used for updating the online models---as \droidevolver uses predicted labels as pseudo-labels, the negative feedback loop is difficult to recover from.
Surprisingly, the drift identification mechanism rejects more correct predictions than it keeps for each test period. We posit that the small app buffer fails to sufficiently represent the true app population, which may in turn lead to the negative feedback loop in the later months. Although much more extreme here, this informational inefficiency is also responsible for the variability we see when using ICEs---different dataset splits may be more or less representative of the true distribution and result in better or worse accuracy, a phenomenon that is mitigated by using a CCE. These limitations, among others, have recently been explored in concurrent work by \citet{kan2021deplusplus}.
}

\revised{
\subsection{Beyond Android Malware and SVMs}
\label{sec:beyond-android}
}

\revised{While \tool and conformal evaluation are agnostic to the underlying classifier and feature space, we have so far focused on detecting Android malware with a linear SVM. Here we demonstrate the performance beyond this setting. To simplify the axes of comparison, we apply an ICE to each setting, using credibility p-values and random search for threshold calibration with the same constraints as before.
\paragraph{Windows PE malware with GBDT} We take examples from the \ember v2 dataset~\citep{anderson2018ember} spanning 2017, containing 47,888 benign and 69,202 malicious executables (labeled as having 40+ VirusTotal AV detections).
The feature space contains a diverse set of features which can be categorized as either parsed features (\eg header information), histograms (\eg byte-value histograms), and printable strings (\eg URL frequency). As the underlying classifier, we use gradient boosted decision trees (GBDT)~\citep{friedman2001greedy} as in \citet{anderson2018ember}, and for the NCM we use the output probability for the predicted class, negated for positive predictions. We train on executables from the first five months and test on the remaining.
\paragraph{PDF malware with RF} We use examples from the \hidost dataset~\citep{srndic2016hidost} spanning five weeks in Aug--Sep 2012, consisting of 181,792 benign and 7,163 malicious files (labeled as having 5+ VirusTotal AV detections).
The feature space is created by statically parsing the PDF files to extract structural paths in the PDF hierarchy that map to boolean or numeric feature values, such as the presence of certain PDF objects or metadata such as the number of pages. As the underlying classifier we use a random forest (RF) classifier following~\citet{srndic2016hidost}. As the NCM we use the proportion of decision trees that disagree with the prediction of the ensemble (as illustrated in \autoref{fig:ncm-rf}). Interestingly, a major contribution of the Hidost feature space in contrast to prior approaches~\citep[\eg][]{srndic13pdf} is that similar features are consolidated  in order to be \textit{more robust to drift}. This means the distribution should be relatively stationary compared to the Android dataset and will allow us to test whether \tool is able to make effective decisions on prediction quality when drift is less severe.
}

\begin{figure}[!tb]
    \centering
        \begin{subfigure}[t]{0.48\columnwidth}
        \centering
        \includegraphics[height=1.5in]{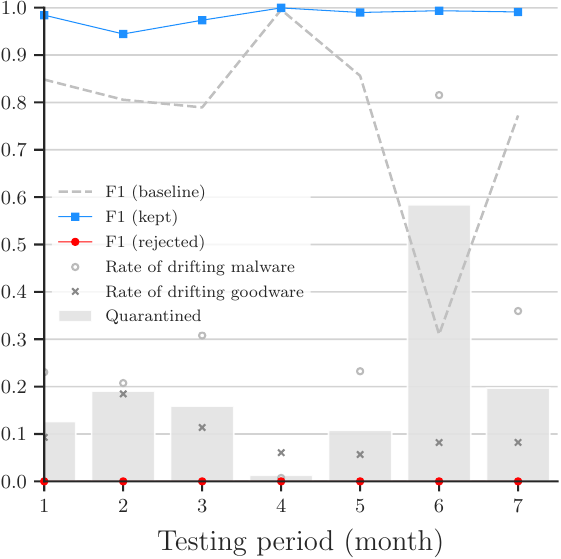}
        \caption{ICE, credibility}
        \label{fig:pe-cred}
    \end{subfigure}
    \hfill
    \begin{subfigure}[t]{0.48\columnwidth}
        \centering
        \includegraphics[height=1.5in]{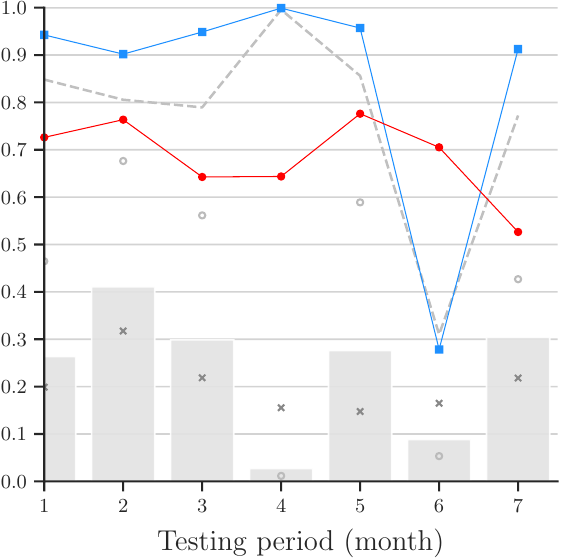}
        \caption{ICE, probabilities}
        \label{fig:pe-proba}
    \end{subfigure}

    \caption{\fone-Score, \ember Windows PE malware~\citep{anderson2018ember} and GBDT.}
    \label{fig:pe-ember-results}
\end{figure}

\revised{
Note that we are unable to find authoritative measurements for the expected class balance for PE and PDF malware in the wild as we are for Android malware, so we defer to the class balance in the original datasets. This may result in a slight spatial bias if the class balance is unrealistic~\citep{pendlebury2019tesseract}, however all approaches will be affected equally. Additionally, we can here examine whether the class balance affects the ability for \tool to identify low quality predictions.
\paragraph{Results} The results for Windows PE malware (\autoref{fig:pe-ember-results}) are consistent with those on Android data. \tool outperforms probabilities alone which tend to reject many otherwise correct predictions. In particular, a large spike in drifting malware affects month six which probabilities are unable to cope with, while \tool raises the rate of rejections accordingly without making any additional errors.
}

\revised{
As noted earlier, the PDF dataset gives us the opportunity to evaluate \tool on a relatively stationary distribution. As expected, thresholding using probabilities is much more effective than it is in a drifting setting, however it under-rejects compared to \tool, which is able to find thresholds that push the \fone of kept predictions to 1.0 while rejecting almost entirely incorrect predictions. Exceptions to this are months one and nine, in which a small quantity of true positive predictions are rejected. However this is anomalous (\ie it does not continue as drift increases) and could be mitigated by calibrating with a constraint on the \fone of rejected samples rather than the \fone of kept examples alone. From this we conclude that \tool is useful for maximizing the potential of a high-quality robust classifier, and does not rely on relatively severe drift---as present in the Android dataset---to detect low quality decisions. To this end \tool can be combined with robust feature spaces which is an orthogonal direction to combating concept drift.
}

\revised{
As an additional result we observe that \tool outperforms \cpreject for both domains, with more details reported in \Cref{app:cpreject}.
}

\begin{figure}[!tb]
    \centering
        \begin{subfigure}[t]{0.48\columnwidth}
        \centering
        \includegraphics[height=1.5in]{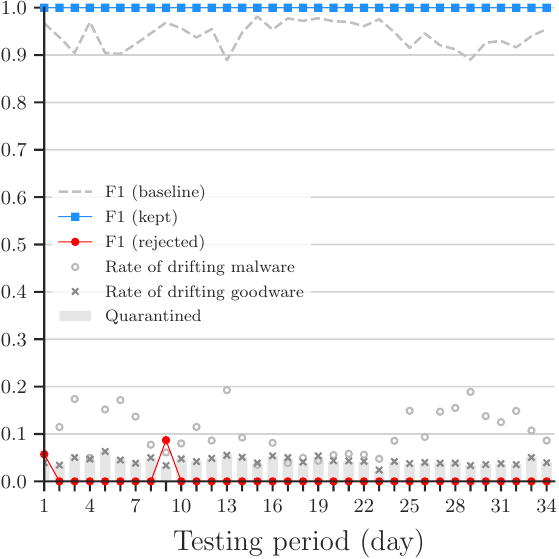}
        \caption{ICE, credibility}
        \label{fig:pdf-cred}
    \end{subfigure}
    \hfill
    \begin{subfigure}[t]{0.48\columnwidth}
        \centering
        \includegraphics[height=1.5in]{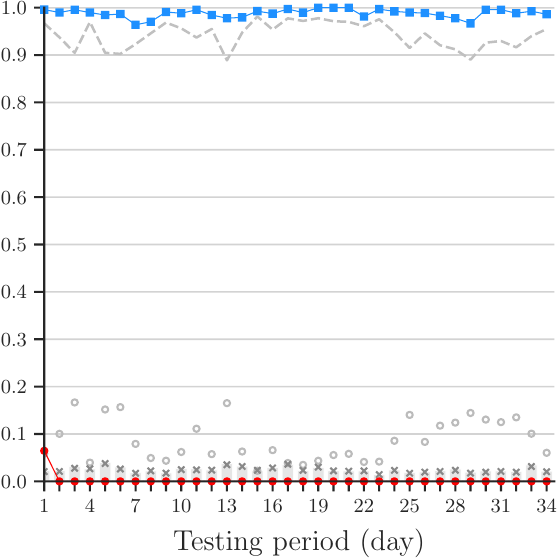}
        \caption{ICE, probabilities}
        \label{fig:pdf-proba}
    \end{subfigure}

    \caption{\fone-Score, \hidost PDF malware~\citep{srndic2016hidost} and RBF SVM.}
    \label{fig:pdf-hidost-results}
\end{figure}

\section{Operational Considerations}
\label{sec:discussion}

Here we discuss some actionable points regarding the use of conformal evaluation and \tool.

\paragraph{\tool in a Detection Pipeline} \tool has particular applications in detection tasks where there is a high cost of individual False Positives, (\eg spam~\cite{nilizadeh17spam}, malware~\cite{srndic13pdf,arp2014drebin,curtsinger11zozzle}, fake accounts~\cite{cao2012sybilrank,boshmaf2015integro}). In these cases, it may be preferable to avoid taking a decision on low-confidence predictions or, where a gradated response is possible, diverting rejected examples towards alternative remediation actions. Consider an example in the fake accounts setting: owners of accounts in the set of rejected positive predictions can be asked to solve a CAPTCHA on their next login (a relatively easy check to pass) while the owners of accounts in the set of kept positive predictions can be asked to submit proof of their identity. Increasing rejection rates signal a performance degradation of the underlying classifier without immediately submitting to the errors it produces, giving engineers more time to iterate and remediate.

\paragraph{Operational Recommendations} Based on our empirical evaluation (\Cref{sec:eval}), we make the following recommendations for \tool deployments:

\begin{itemize}[leftmargin=*]%
	\item \tool is agnostic to the underlying learning algorithm, but the quality of the rejection relies on the suitability of the NCM. Some examples of possible NCMs for different types of classifiers are described in~\Cref{fig:ncm-examples}.
	\item Using an ICE or CCE is preferred over TCE due to their computational efficiency, and is preferred over approx-TCE due to approx-TCE's reliance on assumptions that may not universally hold.
	\item ICEs are relatively fast and lightweight and excel when resources are limited. CCEs make rejections with higher confidence but at a higher computational cost.
	\item Thresholding with credibility alone is sufficient to achieve high quality prediction across all conformal evaluators. While confidence can improve the stability of an ICE (\cref{sec:credconfeval}), it requires greater calibration time.
	\item Random search is preferred over grid search as it finds similarly effective thresholds at significantly lower cost.
	\item Rising rejection rates should be interpreted as a signal that the underlying model is degrading. This signal can be used to trigger model retraining or other remediation strategies.
	\item \revised{Guidance on tuning calibration thresholds and an example of using alternative constraints is presented in~\Cref{app:thresholds}.}
\end{itemize}

\begin{table}
\caption{Performance of optimal thresholds discovered using a full grid search vs.\ random search. Random search discovers thresholds equivalent to the full grid search but with two orders of magnitude fewer trials~(\Cref{sec:search-eval}).}
    \footnotesize
    \centering
\begin{tabular}{lrrrrrr}
  \toprule
   & {\sc FPs}& {\sc FNs} & {\sc Prec.} & {\sc Rec.} & {\sc \fone} & {\sc \#Trials} \\
  \midrule
No rejection  &  3,529 & 19,486 &      0.98 &      0.92 &      0.95 &       N/A \\
Full grid 	  &  2,187 &      0 &      0.99 &      1.00 &      0.99 & 1,317,520 \\
Random    	  &  3,259 &      0 &      0.98 &      1.00 &      0.99 &    10,000 \\
  \bottomrule
\end{tabular}
\label{tab:search}
\end{table}

\section{Related Work}

Conformal evaluation is based on conformal prediction theory, a mechanism for producing predictions that are correct with some guaranteed confidence~\cite{shafer2008tutorial}. Additionally, the ICE and CCE are inspired by inductive~\cite{randomworld, vovk13inductive, papadopoulos08inductive} and cross-conformal predictors~\cite{vovk2018ccp}, respectively. However, conformal prediction is intended to be used exclusively in settings where the exchangeability assumption holds which makes it unsuitable for adversarial contexts such as malware classification. In this regard, we are the first to `join the dots' between the conformal prediction of ~\citet{randomworld} and the conformal evaluation of ~\citet{jordaney2017transcend} and show how the violation of conformal prediction's assumptions is detected and exploited by \transcend to detect concept drift.

That work introduced the concept of \textit{conformal evaluation} based on conformal prediction theory and the use of p-values for calibrating and enforcing a rejection strategy for malware classification. However the evaluation artificially simulated concept drift by merging malware datasets which introduced experimental bias~\cite{pendlebury2019tesseract, arp2020dodo}~(\Cref{sec:eval}). In our experiments we sample from a single repository of applications and perform a temporal evaluation to simulate natural concept drift caused by the real evolution of malicious Android apps. Additionally, the role of confidence in thresholding was unclear, and the use of exhaustive grid search to find thresholds was suboptimal compared to our random search. Most significantly, the TCE originally employed was not practical for real-world deployments, which we rectify by proposing the ICE and CCE.

CADE~\citep{yang2021cade} focuses on \textit{explaining} drift and relies on contrastive learning to perform a distance-based feature transformation which results in more homogenous class clusters relative to which outliers are easier to detect. 
They compare against using conformal evaluation as an active learning query strategy using an arbitrary NCM and \textit{without} \transcend thresholding.
We believe CADE and \tool are orthogonal and that such feature transformations can be used to devise stronger NCMs---we leave the development of optimal NCMs for active learning as future work.

Other works have explored alternative solutions to tackling concept drift. \revised{As described in~\Cref{sec:prior-methods}, \droidevolver~\citep{xu2019edvolver, kan2021deplusplus} is a malware detection system motivated by \transcend that identifies drifting examples based on disagreements between models in an ensemble. As models degrade, the examples identified as drifting are used to update the models in an online fashion. However, we find the drift identification mechanism is inferior to \tool and leads to a negative feedback loop.} Other solutions solely adapt to concept drift without using rejection:  \text{DroidOL}~\cite{droidol} and \text{Casandra}~\cite{casandra} use online learning to continually retrain the models, with API call graphs as features. Like all online-trained neural networks, these are susceptible to catastrophic forgetting~\cite{french02catastrophic}, where performance degrades on older examples as the model attempts to adapt to the new distribution. \citet{pendlebury2019tesseract} present a comparison of different strategies for combating concept drift, including rejection, incremental retraining, and online learning, illustrating the advantages and disadvantages of each.
Semi-supervised techniques may be used to reduce the labeling burden of such strategies as drift increases, as recently demonstrated for intrusion detection by \citet{andresini2021insomnia}. 

The related task of detecting \textit{adversarial examples}~\cite{szegedy2014intriguing, biggio2018wild, pierazzi2020problemspace} is addressed by  \citet{sotgiu20deeprejection}, who propose a rejection strategy for neural network-based classifiers that identifies anomalies in an input's latent feature representation at different layers of the neural network. Additionally, \citet{papernot18deepknn} combine a conformal predictor with a $k$-Nearest Neighbor algorithm to identify low-quality predictions that are indicative of adversarial inputs. However, both methods are restricted to deep learning-based image classification.

\section{Conclusion}

We provide a thorough formal treatment of \transcend which acts as
the \textit{missing link} between conformal prediction and conformal
evaluation. We propose \tool, a superset of the original framework which includes
novel conformal evaluators that match or
surpass the original performance while significantly decreasing
the computational cost. We show \tool outperforms the
existing state-of-the-art approaches while generalizing across different
malware domains and exploring realistic operational settings. 

We envision these improvements will enable researchers and practitioners
alike to make use of conformal evaluation to build
rejection strategies to improve their security detection
pipelines. To this end, we release our implementation of \tool,
making \transcend and conformal evaluation available to the community for the first time.

\section*{Acknowledgements}
This research has been partially sponsored by the UK EP/P009301/1 EPSRC research grant.

{\footnotesize
\bibliography{bibliography}

\begin{thebibliography}{53}
\providecommand{\natexlab}[1]{#1}
\providecommand{\url}[1]{\texttt{#1}}
\expandafter\ifx\csname urlstyle\endcsname\relax
  \providecommand{\doi}[1]{doi: #1}\else
  \providecommand{\doi}{doi: \begingroup \urlstyle{rm}\Url}\fi

\bibitem[Allix et~al.(2015)Allix, Bissyand{\'{e}}, Klein, and
  Traon]{allix15relevant}
K.~Allix, T.~F. Bissyand{\'{e}}, J.~Klein, and Y.~L. Traon.
\newblock Are your training datasets yet relevant? - an investigation into the
  importance of timeline in machine learning-based malware detection.
\newblock In \emph{International Symposium on Engineering Secure Software and
  Systems (ESSoS)}, 2015.

\bibitem[Allix et~al.(2016)Allix, Bissyand{\'e}, Klein, and Le~Traon]{androzoo}
K.~Allix, T.~F. Bissyand{\'e}, J.~Klein, and Y.~Le~Traon.
\newblock Androzoo: Collecting millions of android apps for the research
  community.
\newblock In \emph{{ACM} International Conference on Mining Software
  Repositories ({MSR})}, 2016.

\bibitem[Anderson and Roth(2018)]{anderson2018ember}
H.~S. Anderson and P.~Roth.
\newblock {EMBER:} an open dataset for training static {PE} malware machine
  learning models.
\newblock \emph{CoRR}, abs/1804.04637, 2018.

\bibitem[Anderson et~al.(2018)Anderson, Kharkar, Filar, Evans, and
  Roth]{anderson2018pdfgan}
H.~S. Anderson, A.~Kharkar, B.~Filar, D.~Evans, and P.~Roth.
\newblock Learning to evade static {PE} machine learning malware models via
  reinforcement learning.
\newblock \emph{CoRR}, abs/1801.08917, 2018.

\bibitem[Andresini et~al.(2021)Andresini, Pendlebury, Pierazzi, Loglisci,
  Appice, and Cavallaro]{andresini2021insomnia}
G.~Andresini, F.~Pendlebury, F.~Pierazzi, C.~Loglisci, A.~Appice, and
  L.~Cavallaro.
\newblock {INSOMNIA:} towards concept-drift robustness in network intrusion
  detection.
\newblock In \emph{{ACM} Workshop on Artificial Intelligence and Security
  ({AISec})}, 2021.

\bibitem[Arp et~al.(2014)Arp, Spreitzenbarth, Hubner, Gascon, and
  Rieck]{arp2014drebin}
D.~Arp, M.~Spreitzenbarth, M.~Hubner, H.~Gascon, and K.~Rieck.
\newblock {DREBIN:} effective and explainable detection of android malware in
  your pocket.
\newblock In \emph{Network and Distributed System Security Symposium ({NDSS})},
  2014.

\bibitem[Arp et~al.(2022)Arp, Quiring, Pendlebury, Warnecke, Pierazzi,
  Wressnegger, Cavallaro, and Rieck]{arp2020dodo}
D.~Arp, E.~Quiring, F.~Pendlebury, A.~Warnecke, F.~Pierazzi, C.~Wressnegger,
  L.~Cavallaro, and K.~Rieck.
\newblock Dos and don'ts of machine learning in computer security.
\newblock In \emph{{USENIX} Security Symposium}, 2022.

\bibitem[Bartlett and Wegkamp(2008)]{bartlett08rejection}
P.~L. Bartlett and M.~H. Wegkamp.
\newblock Classification with a reject option using a hinge loss.
\newblock \emph{Journal of Machine Learning Research ({JMLR})}, 2008.

\bibitem[Bellman(2015)]{bellman61curse}
R.~Bellman.
\newblock \emph{Adaptive Control Processes - {A} Guided Tour (Reprint from
  1961)}.
\newblock Princeton University Press, 2015.

\bibitem[Bergstra and Bengio(2012)]{bergstra12search}
J.~Bergstra and Y.~Bengio.
\newblock Random search for hyper-parameter optimization.
\newblock \emph{Journal of Machine Learning Research ({JMLR})}, 2012.

\bibitem[Biggio and Roli(2018)]{biggio2018wild}
B.~Biggio and F.~Roli.
\newblock Wild patterns: Ten years after the rise of adversarial machine
  learning.
\newblock \emph{Pattern Recognition}, 2018.

\bibitem[Bishop(2007)]{bishop}
C.~M. Bishop.
\newblock \emph{Pattern recognition and machine learning, 5th Edition}.
\newblock Information science and statistics. Springer, 2007.

\bibitem[Boshmaf et~al.(2015)Boshmaf, Logothetis, Siganos, Ler{\'{\i}}a,
  Lorenzo, Ripeanu, and Beznosov]{boshmaf2015integro}
Y.~Boshmaf, D.~Logothetis, G.~Siganos, J.~Ler{\'{\i}}a, J.~Lorenzo, M.~Ripeanu,
  and K.~Beznosov.
\newblock Integro: Leveraging victim prediction for robust fake account
  detection in {OSNs}.
\newblock In \emph{Network and Distributed System Security Symposium ({NDSS})},
  2015.

\bibitem[Cao et~al.(2012)Cao, Sirivianos, Yang, and
  Pregueiro]{cao2012sybilrank}
Q.~Cao, M.~Sirivianos, X.~Yang, and T.~Pregueiro.
\newblock Aiding the detection of fake accounts in large scale social online
  services.
\newblock In \emph{{USENIX} Symposium on Networked Systems Design and
  Implementation ({NSDI})}, 2012.

\bibitem[Curtsinger et~al.(2011)Curtsinger, Livshits, Zorn, and
  Seifert]{curtsinger11zozzle}
C.~Curtsinger, B.~Livshits, B.~G. Zorn, and C.~Seifert.
\newblock {ZOZZLE:} fast and precise in-browser javascript malware detection.
\newblock In \emph{{USENIX} Security Symposium}, 2011.

\bibitem[Dash et~al.(2016)Dash, Suarez{-}Tangil, Khan, Tam, Ahmadi, Kinder, and
  Cavallaro]{dash2016droidscribe}
S.~K. Dash, G.~Suarez{-}Tangil, S.~J. Khan, K.~Tam, M.~Ahmadi, J.~Kinder, and
  L.~Cavallaro.
\newblock Droidscribe: Classifying android malware based on runtime behavior.
\newblock In \emph{{IEEE} Security and Privacy Workshops ({SPW})}, 2016.

\bibitem[Edunov et~al.(2018)Edunov, Ott, Auli, and Grangier]{backtranslate}
S.~Edunov, M.~Ott, M.~Auli, and D.~Grangier.
\newblock Understanding back-translation at scale.
\newblock In \emph{Conference on Empirical Methods in Natural Language
  Processing (EMNLP)}, 2018.

\bibitem[French and Chater(2002)]{french02catastrophic}
R.~M. French and N.~Chater.
\newblock Using noise to compute error surfaces in connectionist networks: {A}
  novel means of reducing catastrophic forgetting.
\newblock \emph{Neural Computation}, 2002.

\bibitem[Friedman(2001)]{friedman2001greedy}
J.~H. Friedman.
\newblock Greedy function approximation: a gradient boosting machine.
\newblock \emph{Annals of Statistics}, 2001.

\bibitem[Jordaney et~al.(2017)Jordaney, Sharad, Dash, Wang, Papini,
  Nouretdinov, and Cavallaro]{jordaney2017transcend}
R.~Jordaney, K.~Sharad, S.~K. Dash, Z.~Wang, D.~Papini, I.~Nouretdinov, and
  L.~Cavallaro.
\newblock Transcend: Detecting concept drift in malware classification models.
\newblock In \emph{{USENIX} Security Symposium}, 2017.

\bibitem[Kan et~al.(2021)Kan, Pendlebury, Pierazzi, and
  Cavallaro]{kan2021deplusplus}
Z.~Kan, F.~Pendlebury, F.~Pierazzi, and L.~Cavallaro.
\newblock Investigating labelless drift adaptation for malware detection.
\newblock In \emph{{ACM} Workshop on Artificial Intelligence and Security
  ({AISec})}, 2021.

\bibitem[Kantchelian et~al.(2015)Kantchelian, Tschantz, Afroz, Miller, Shankar,
  Bachwani, Joseph, and Tygar]{kantchelian2015regression}
A.~Kantchelian, M.~C. Tschantz, S.~Afroz, B.~Miller, V.~Shankar, R.~Bachwani,
  A.~D. Joseph, and J.~D. Tygar.
\newblock Better malware ground truth: Techniques for weighting anti-virus
  vendor labels.
\newblock In \emph{{ACM} Workshop on Artificial Intelligence and Security
  ({AISec})}, 2015.

\bibitem[Krizhevsky et~al.(2017)Krizhevsky, Sutskever, and Hinton]{alexnet}
A.~Krizhevsky, I.~Sutskever, and G.~E. Hinton.
\newblock Imagenet classification with deep convolutional neural networks.
\newblock \emph{Commun. {ACM}}, 2017.

\bibitem[Kullback and Leibler(1951)]{kullback1951}
S.~Kullback and R.~A. Leibler.
\newblock On information and sufficiency.
\newblock \emph{Annals of Mathematical Statistics}, 1951.

\bibitem[Lindorfer et~al.(2015)Lindorfer, Neugschwandtner, and Platzer]{marvin}
M.~Lindorfer, M.~Neugschwandtner, and C.~Platzer.
\newblock {MARVIN:} efficient and comprehensive mobile app classification
  through static and dynamic analysis.
\newblock In \emph{{IEEE} Annual Computer Software and Applications Conference
  ({COMPSAC})}, 2015.

\bibitem[Linusson et~al.(2018)Linusson, Johansson, Bostr{\"{o}}m, and
  L{\"{o}}fstr{\"{o}}m]{linusson2018pakdd}
H.~Linusson, U.~Johansson, H.~Bostr{\"{o}}m, and T.~L{\"{o}}fstr{\"{o}}m.
\newblock Classification with reject option using conformal prediction.
\newblock In \emph{Pacific-Asia Conference on Knowledge Discovery and Data
  Mining ({PAKDD})}. Springer, 2018.

\bibitem[Miller et~al.(2016)Miller, Kantchelian, Tschantz, Afroz, Bachwani,
  Faizullabhoy, Huang, Shankar, Wu, Yiu, Joseph, and Tygar]{miller16dimva}
B.~Miller, A.~Kantchelian, M.~C. Tschantz, S.~Afroz, R.~Bachwani,
  R.~Faizullabhoy, L.~Huang, V.~Shankar, T.~Wu, G.~Yiu, A.~D. Joseph, and J.~D.
  Tygar.
\newblock Reviewer integration and performance measurement for malware
  detection.
\newblock In \emph{Conference on Detection of Intrusions and Malware {\&}
  Vulnerability Assessment ({DIMVA})}, 2016.

\bibitem[Moreno{-}Torres et~al.(2012)Moreno{-}Torres, Raeder,
  Ala{\'{\i}}z{-}Rodr{\'{\i}}guez, Chawla, and Herrera]{datasetshift}
J.~G. Moreno{-}Torres, T.~Raeder, R.~Ala{\'{\i}}z{-}Rodr{\'{\i}}guez, N.~V.
  Chawla, and F.~Herrera.
\newblock A unifying view on dataset shift in classification.
\newblock \emph{Pattern Recognition}, 2012.

\bibitem[Narayanan et~al.(2016)Narayanan, Liu, Chen, and Liu]{droidol}
A.~Narayanan, Y.~Liu, L.~Chen, and J.~Liu.
\newblock Adaptive and scalable android malware detection through online
  learning.
\newblock In \emph{International Joint Conference on Neural Network ({IJCNN})},
  2016.

\bibitem[Narayanan et~al.(2017)Narayanan, Chandramohan, Chen, and
  Liu]{casandra}
A.~Narayanan, M.~Chandramohan, L.~Chen, and Y.~Liu.
\newblock Context-aware, adaptive, and scalable android malware detection
  through online learning.
\newblock \emph{{IEEE} Transactions on Emerging Topics in Computational
  Intelligence ({TETCI})}, 2017.

\bibitem[Nilizadeh et~al.(2017)Nilizadeh, Labreche, Sedighian, Zand, Fernandez,
  Kruegel, Stringhini, and Vigna]{nilizadeh17spam}
S.~Nilizadeh, F.~Labreche, A.~Sedighian, A.~Zand, J.~M. Fernandez, C.~Kruegel,
  G.~Stringhini, and G.~Vigna.
\newblock {POISED:} spotting twitter spam off the beaten paths.
\newblock In \emph{{ACM} Conference on Computer and Communications Security
  ({CCS})}, 2017.

\bibitem[Papadopoulos(2008)]{papadopoulos08inductive}
H.~Papadopoulos.
\newblock Inductive conformal prediction: Theory and application to neural
  networks.
\newblock In \emph{Tools in Artificial Intelligence}. 2008.

\bibitem[Papernot and McDaniel(2018)]{papernot18deepknn}
N.~Papernot and P.~D. McDaniel.
\newblock Deep k-nearest neighbors: Towards confident, interpretable and robust
  deep learning.
\newblock \emph{CoRR}, abs/1803.04765, 2018.

\bibitem[Pendlebury(2021)]{pendlebury2021hostile}
F.~Pendlebury.
\newblock \emph{Machine Learning for Security in Hostile Environments}.
\newblock PhD thesis, University of London, 2021.

\bibitem[Pendlebury et~al.(2019)Pendlebury, Pierazzi, Jordaney, Kinder, and
  Cavallaro]{pendlebury2019tesseract}
F.~Pendlebury, F.~Pierazzi, R.~Jordaney, J.~Kinder, and L.~Cavallaro.
\newblock {TESSERACT:} eliminating experimental bias in malware classification
  across space and time.
\newblock In \emph{{USENIX} Security Symposium}, 2019.

\bibitem[Pierazzi et~al.(2020)Pierazzi, Pendlebury, Cortellazzi, and
  Cavallaro]{pierazzi2020problemspace}
F.~Pierazzi, F.~Pendlebury, J.~Cortellazzi, and L.~Cavallaro.
\newblock Intriguing properties of adversarial {ML} attacks in the problem
  space.
\newblock In \emph{{IEEE} Symposium on Security and Privacy ({S\&P})}, 2020.

\bibitem[Plato(c. 385–370 BC)]{plato}
Plato.
\newblock \emph{The Symposium}.
\newblock c. 385–370 BC.
\newblock Penguin Classics edition published 1999, translated by Christopher
  Gill.

\bibitem[Shafer and Vovk(2008)]{shafer2008tutorial}
G.~Shafer and V.~Vovk.
\newblock A tutorial on conformal prediction.
\newblock \emph{Journal of Machine Learning Research ({JMLR})}, 2008.

\bibitem[Sotgiu et~al.(2020)Sotgiu, Demontis, Melis, Biggio, Fumera, Feng, and
  Roli]{sotgiu20deeprejection}
A.~Sotgiu, A.~Demontis, M.~Melis, B.~Biggio, G.~Fumera, X.~Feng, and F.~Roli.
\newblock Deep neural rejection against adversarial examples.
\newblock \emph{{EURASIP} Journal on Information Security}, 2020.

\bibitem[Srndic and Laskov(2013)]{srndic13pdf}
N.~Srndic and P.~Laskov.
\newblock Detection of malicious {PDF} files based on hierarchical document
  structure.
\newblock In \emph{Network and Distributed System Security Symposium ({NDSS})},
  2013.

\bibitem[Srndic and Laskov(2016)]{srndic2016hidost}
N.~Srndic and P.~Laskov.
\newblock Hidost: a static machine-learning-based detector of malicious files.
\newblock \emph{{EURASIP} Journal on Information Security}, 2016.

\bibitem[Suarez{-}Tangil et~al.(2014)Suarez{-}Tangil, Tapiador, Peris{-}Lopez,
  and Al{\'{\i}}s]{guillermo2014dendroid}
G.~Suarez{-}Tangil, J.~E. Tapiador, P.~Peris{-}Lopez, and J.~B. Al{\'{\i}}s.
\newblock Dendroid: {A} text mining approach to analyzing and classifying code
  structures in android malware families.
\newblock \emph{Expert Systems With Applications}, 2014.

\bibitem[Suarez{-}Tangil et~al.(2017)Suarez{-}Tangil, Dash, Ahmadi, Kinder,
  Giacinto, and Cavallaro]{guillermo2017droidsieve}
G.~Suarez{-}Tangil, S.~K. Dash, M.~Ahmadi, J.~Kinder, G.~Giacinto, and
  L.~Cavallaro.
\newblock Droidsieve: Fast and accurate classification of obfuscated android
  malware.
\newblock In \emph{{ACM} Conference on Data and Applications Security and
  Privacy (CODASPY)}, 2017.

\bibitem[Szegedy et~al.(2014)Szegedy, Zaremba, Sutskever, Bruna, Erhan,
  Goodfellow, and Fergus]{szegedy2014intriguing}
C.~Szegedy, W.~Zaremba, I.~Sutskever, J.~Bruna, D.~Erhan, I.~J. Goodfellow, and
  R.~Fergus.
\newblock Intriguing properties of neural networks.
\newblock In \emph{{ICLR} (Poster)}, 2014.

\bibitem[Tong et~al.(2019)Tong, Li, Hajaj, Xiao, Zhang, and
  Vorobeychik]{liang2019conserved}
L.~Tong, B.~Li, C.~Hajaj, C.~Xiao, N.~Zhang, and Y.~Vorobeychik.
\newblock Improving robustness of {ML} classifiers against realizable evasion
  attacks using conserved features.
\newblock In \emph{{USENIX} Security Symposium}, 2019.

\bibitem[Vovk(2013)]{vovk13inductive}
V.~Vovk.
\newblock Conditional validity of inductive conformal predictors.
\newblock \emph{Journal of Machine Learning Research ({JMLR})}, 2013.

\bibitem[Vovk et~al.(2010)Vovk, Gammerman, and Shafer]{randomworld}
V.~Vovk, A.~Gammerman, and G.~Shafer.
\newblock \emph{Algorithmic learning in a random world}.
\newblock Springer-verlag New York Inc., 2010.

\bibitem[Vovk et~al.(2018)Vovk, Nouretdinov, Manokhin, and
  Gammerman]{vovk2018ccp}
V.~Vovk, I.~Nouretdinov, V.~Manokhin, and A.~Gammerman.
\newblock Cross-conformal predictive distributions.
\newblock In \emph{{Workshop on Conformal Prediction and its Applications
  (COPA)}}, 2018.

\bibitem[Xi et~al.(2019)Xi, Yang, Xiao, Yao, Xiong, Xu, Wang, Gao, Liu, Xu, and
  Lu]{xi2019deepintent}
S.~Xi, S.~Yang, X.~Xiao, Y.~Yao, Y.~Xiong, F.~Xu, H.~Wang, P.~Gao, Z.~Liu,
  F.~Xu, and J.~Lu.
\newblock Deepintent: Deep icon-behavior learning for detecting
  intention-behavior discrepancy in mobile apps.
\newblock In \emph{{ACM} Conference on Computer and Communications Security
  ({CCS})}, 2019.

\bibitem[Xu et~al.(2019)Xu, Li, Deng, Chen, and Xu]{xu2019edvolver}
K.~Xu, Y.~Li, R.~H. Deng, K.~Chen, and J.~Xu.
\newblock Droidevolver: Self-evolving android malware detection system.
\newblock In \emph{{IEEE} European Symposium on Security and Privacy
  ({EuroS\&P})}, 2019.

\bibitem[Yang et~al.(2021)Yang, Guo, Hao, Ciptadi, Ahmadzadeh, Xing, and
  Wang]{yang2021cade}
L.~Yang, W.~Guo, Q.~Hao, A.~Ciptadi, A.~Ahmadzadeh, X.~Xing, and G.~Wang.
\newblock {CADE:} detecting and explaining concept drift samples for security
  applications.
\newblock In \emph{{USENIX} Security Symposium}, 2021.

\bibitem[Yang et~al.(2015)Yang, Xiao, Andow, Li, Xie, and
  Enck]{yang2015appcontext}
W.~Yang, X.~Xiao, B.~Andow, S.~Li, T.~Xie, and W.~Enck.
\newblock Appcontext: Differentiating malicious and benign mobile app behaviors
  using context.
\newblock In \emph{International Conference on Software Engineering ({ICSE})},
  2015.

\bibitem[Zhang et~al.(2020)Zhang, Zhang, Zhong, Ding, Cao, Zhang, Zhang, and
  Yang]{zhang2020apigraph}
X.~Zhang, Y.~Zhang, M.~Zhong, D.~Ding, Y.~Cao, Y.~Zhang, M.~Zhang, and M.~Yang.
\newblock {Enhancing State-of-the-Art Classifiers with API Semantics to Detect
  Evolved Android Malware}.
\newblock In \emph{{ACM} Conference on Computer and Communications Security
  ({CCS})}, 2020.

\end{thebibliography}
}

\begin{appendix}

\subsection{Symbol Table}
\label{app:symbol}

\autoref{tab:symbols} reports the major symbols and abbreviations used throughout the paper.

\begin{table}[!tb]
\centering
\caption{Table of symbols and abbreviations.}
\label{tab:symbols}
\begin{tabular}{lp{6cm}}
  \toprule
  {\sc Symbol} & {\sc Description} \\
  \midrule
  \(\mathcal{X}\) & Feature space \(\mathcal{X} \subseteq \mathbb{R}^n\). \\
  \(\mathcal{Y}\) & Label space. \\
  \(z\)			& Example pair \((\bm{x}, y) \in \mathcal{X} \times \mathcal{Y}\). \\
  \(z^*\)			& Previously unseen test example. \\
  \(\hat{y}\)		& Predicted class \(g(z^*)\). \\
  \(a_z\)			& Nonconformity score output by an NCM for \(z\). \\
  \(p_z\)			& Statistical p-value for \(z\). \\
  \(p_z^y\)       & Statistical p-value for \(z\), calculated with respect to class \(y \in \mathcal{Y}\) (used in \textit{label conditional} calculations). \\
  \(\tau_{y}\)    & A rejection threshold \(\tau_y \in [0,1]\) for class \(y \in \mathcal{Y}\). \\
  \(\mathcal{T}\) & The set of all per-class rejection thresholds \(\{\, \tau_y \in [0,1] \mid y \in \mathcal{Y} \,\}\). \\
  \(B\) 			& Bag of examples \(\Lbag z_1, z_2, ... , z_n \Rbag \). \\
  \(d\) 			& Distance function \(d(z, z')\). \\
  \(\hat{z}\)     & Point predictor \(\hat{z}(B)\). \\
  \(A\)			& Nonconformity measure (NCM) usually composed of a distance function and point predictor. \\
  \(S\) 			& Collection of nonconformity scores computed in elements of \(B\), relative to other elements in \(B\), \(S = \Lbag A(B \setminus \Lbag z \Rbag, z) : z \in B \Rbag\). \\
  \(g\) 			& Classifier \(g:\mathcal{X}\longrightarrow \mathcal{Y}\) that assigns object \(\bm{x} \in \mathcal{X}\) to class \(y\in \mathcal{Y}\). Also known as the \emph{decision function}.  \\
  \(\varepsilon\) & Significance level used in conformal prediction to define prediction region with confidence guarantees.\\
    NCM  		& Nonconformity measure. \\
    TCE  		& Transductive Conformal Evaluator. \\
    ICE  		& Inductive Conformal Evaluator. \\
    CCE  		& Cross-Conformal Evaluator. \\
	\bottomrule
\end{tabular}
\end{table}

\begin{figure}[!tb]
    \centering
        \begin{subfigure}[t]{0.48\columnwidth}
        \centering
        \includegraphics[height=1.5in]{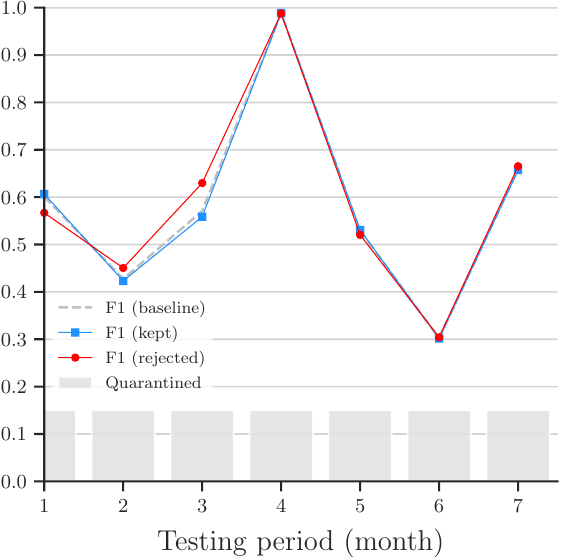}
        \caption{\ember Windows PE malware~\citep{anderson2018ember}}
        \label{fig:pe-pakdd}
    \end{subfigure}
    \hfill
    \begin{subfigure}[t]{0.48\columnwidth}
        \centering
        \includegraphics[height=1.5in]{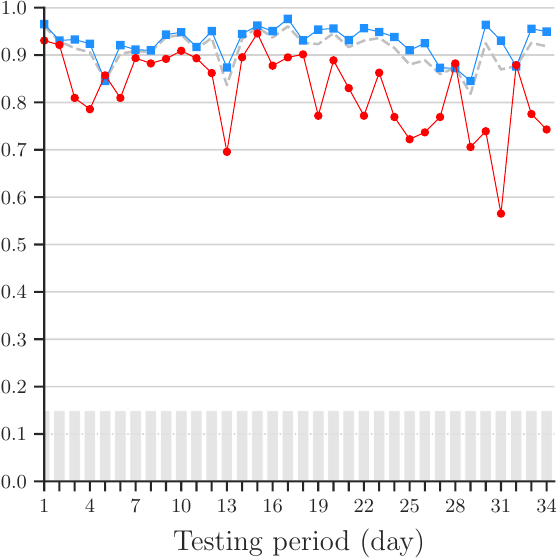}
        \caption{\hidost PDF malware~\citep{srndic2016hidost}}
        \label{fig:pdf-pakdd}
    \end{subfigure}

    \caption{\fone-Score of \cpreject~\citep{linusson2018pakdd} on alternative malware datasets.}
    \label{fig:pakdd-extra}
\end{figure}

\revised{
\subsection{Additional \cpreject Results}
\label{app:cpreject}
In \Cref{sec:beyond-android} we demonstrate how \tool can apply to other classifiers and domains, comparing the performance of an ICE using credibility against using probabilities alone, for both PE and PDF malware~(\Cref{fig:pe-ember-results,fig:pdf-hidost-results}). Here we show in  \autoref{fig:pakdd-extra} additional results to compare against the prior rejection approach \cpreject (we exclude \droidevolver as it is specific to Android malware). Similar to the results on the Android dataset~(\Cref{sec:prior-methods}), the overall ability for \cpreject to distinguish between drifting and non-drifting points is poor on the PE malware dataset. For the PDF malware dataset, which exhibits much less drift, \cpreject is significantly more effective, which supports the hypothesis that it is the violation of conformal prediction's exchangeability assumption which results in the lower performance on the Android and PE datasets. Nevertheless, \tool with credibility (and even probabilities) outperforms \cpreject in this setting also (\cf \Cref{fig:pdf-hidost-results}).
}

\begin{table}[t]
\caption{AUT(\fone, 7$m$) comparing vanilla TCE to our novel conformal evaluators on Windows PE malware data. To be computationally viable, 10\% of the training data was randomly sampled to use for training and calibration.}
    \footnotesize
    \centering
    \begin{tabular}{lP{1cm}P{1.6cm}P{1cm}P{1cm}}
    \toprule
                        &  \textbf{TCE} & \textbf{Approx-TCE} & \textbf{ICE} & \textbf{CCE} \\
    \midrule
    Baseline           & 0.68   & 0.70  & 0.45  & 0.69           \\
    Kept Elements      & 0.97   & 0.97  & 0.94  & 1.00           \\
    Rejected Elements  & 0.00   & 0.00  & 0.00  & 0.21           \\
    \bottomrule
\end{tabular}

\label{tab:full-tce}
\end{table}

\revised{
\subsection{Full Vanilla TCE on \ember Subset}
\label{app:full-tce}
A full scale comparison to the original TCE is not possible due to its computational complexity---recall that one classifier must be trained for each example in the training set. However, it is informative to perform a small-scale experiment as there may be settings where the vanilla TCE \textit{is} viable, and we wish to ensure that there is no significant performance difference between vanilla TCE and our novel conformal evaluators.
}

\revised{
We perform an experiment on the Windows PE malware dataset, where 10\% of the training data is randomly sampled to use for training and calibrating the evaluators (this is the largest subsample we can take given our resource constraints). We choose the PE dataset over the Android dataset due to the high dimensionality of the Android feature space that may cause instability when the number of examples is very low, and over the PDF dataset which is relatively stationary and may make it harder to discern performance differences between the different evaluators. One caveat of this subsampling is the reduced performance of the baseline for the ICE, which is due to the reduced data available to the proper training set, although \tool appears unaffected by this.
}

\revised{
 \Cref{tab:full-tce} summarizes the \fone performance over the seven month-long test periods using the area-under-time (AUT) metric~\cite{pendlebury2019tesseract}. The performance difference between TCE and our evaluators in terms of distinguishing between drifting and non-drifting examples is negligible, shown by the very high AUT of kept elements and very low AUT of rejected elements. That is, there is little to no performance sacrifice when using our evaluators over the vanilla TCE. The overall trends otherwise follow those in our main Android experiments (\cf \autoref{fig:fpr-comparison}).
}

\begin{figure}[!tb]
    \centering
    \includegraphics[height=1.5in]{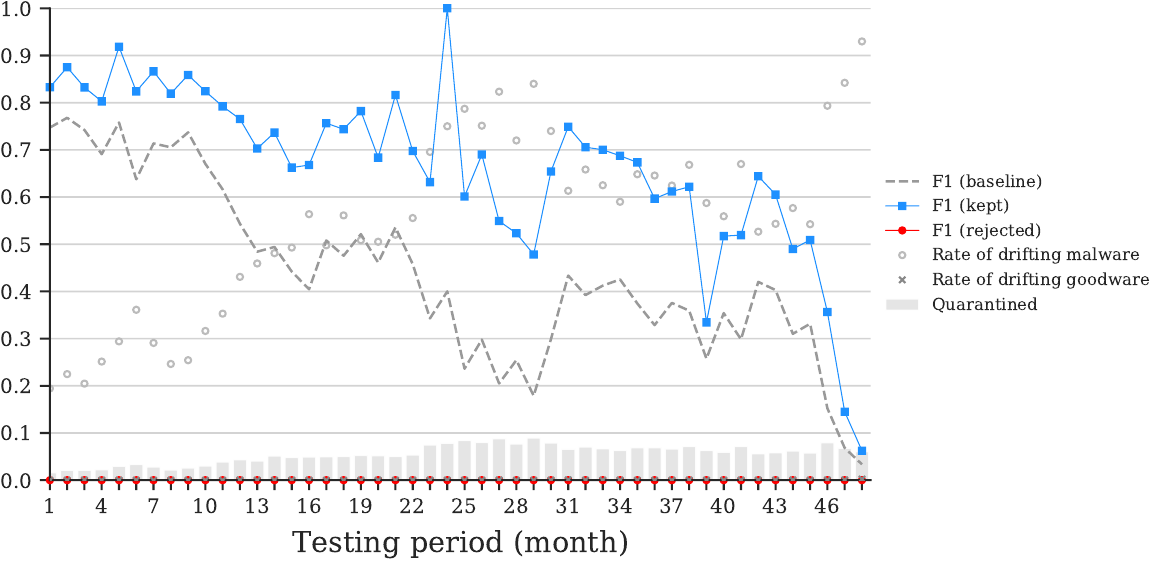}
    \caption{\fone-Score of an ICE optimized to find calibration thresholds that minimize the rejection rate with \fone-Score no less than 0.8. These settings keep the rejection rate low (below 10\%) while sacrificing the \fone performance on kept elements (\cf~\autoref{fig:fpr-f1-ice}).}
    \label{fig:threshold-tuning}
\end{figure}

\begin{figure*}[!t]
    \centering
    \begin{subfigure}[t]{0.3\textwidth}
        \centering
        \includegraphics[width=0.9\columnwidth]{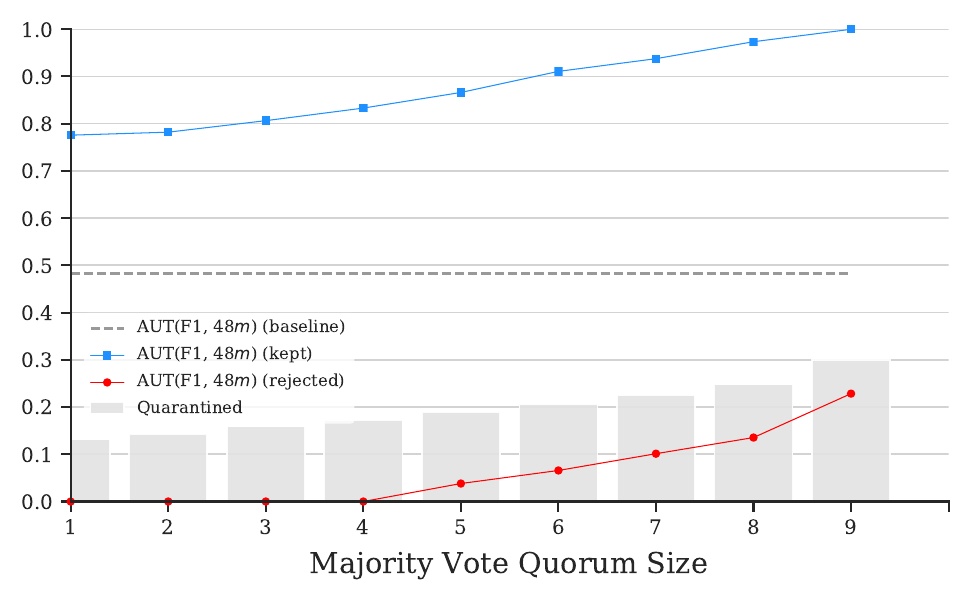}
        \caption{AUT(\fone, 48\textit{m})}
        \label{fig:f1-tce-tuning}
    \end{subfigure}
   	\hfill
    \begin{subfigure}[t]{0.3\textwidth}
        \centering
        \includegraphics[width=0.9\columnwidth]{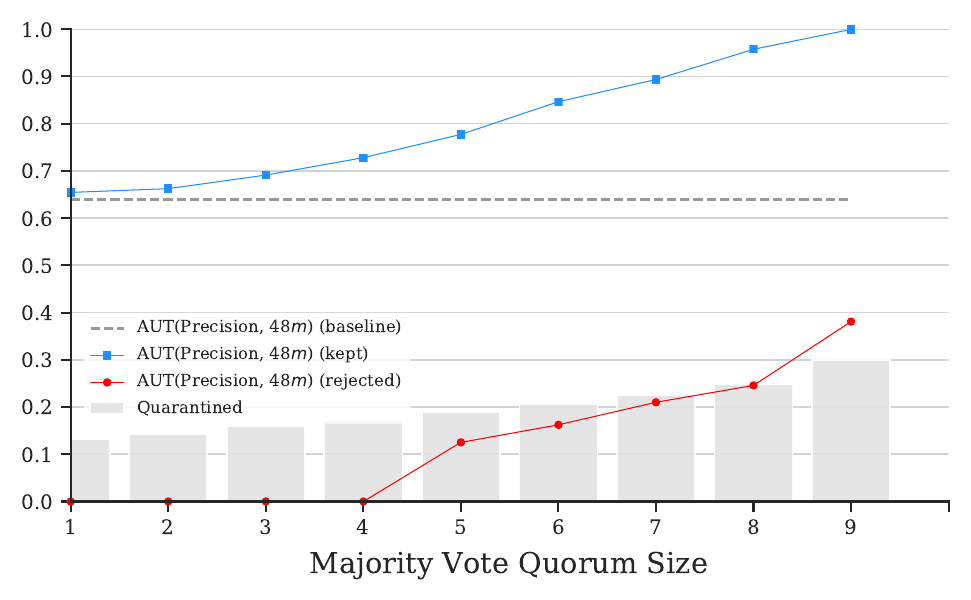}
        \caption{AUT(Precision, 48\textit{m})}
        \label{fig:prec-tce-tuning}
    \end{subfigure}
	\hfill
    \begin{subfigure}[t]{0.3\textwidth}
        \centering
        \includegraphics[width=0.9\columnwidth]{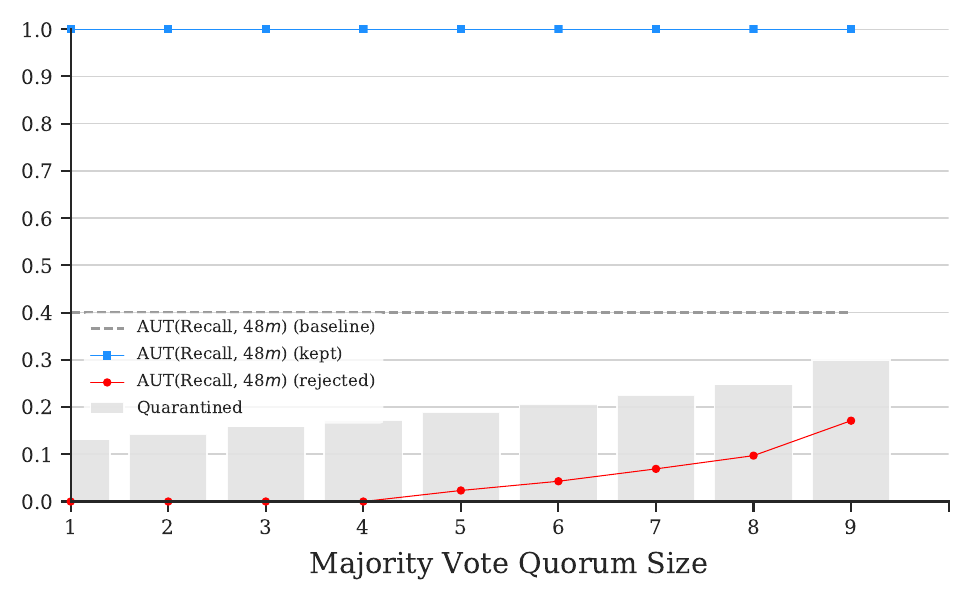}
        \caption{AUT(Recall, 48\textit{m})}
        \label{fig:rec-tce-tuning}
    \end{subfigure}

    \caption{AUT of performance metrics showing the effect of tuning the quorum size $k$ of the majority vote in a CCE.}
    \label{fig:cce-k-tuning}
    \vspace{-2em}
\end{figure*}

\subsection{Analysis of CCE Tuning}
\label{app:cce-tuning}

Here we revisit the majority vote conditions for the CCE applied to the Android malware dataset  in~\Cref{sec:conf-eval}. The size of the quorum for the CCE affects how conservative the CCE is in accepting test examples. \autoref{fig:cce-k-tuning} shows the performance over time summarized using the AUT metric for \fone (a), Precision (b), and Recall (c). Note that \autoref{fig:cce-k-tuning} omits the setting where the majority vote must be unanimous, as the CCE eventually rejects every example---causing \fone, Precision, and Recall to be undefined for kept elements. As more folds of the CCE are required to agree with each other before a decision is accepted, the CCE will reject more elements. If less folds are required, more elements will be accepted. Similarly, the quality of the rejection lessens: more elements are rejected on which the underlying classifier would not have made a mistake. Tuning the majority vote conditions on the calibration set can help find the sweet spot between the performance of kept elements, and the quality---and volume---of rejections.

\revised{
\subsection{Guidance for Choosing Calibration Constraints}
\label{app:thresholds}
In \Cref{sec:search} we formally describe the threshold calibration as an optimization problem in which one metric of interest is maximized or minimized given constraints on another metric. Throughout our evaluation we focus on maximizing the \fone of kept elements, while keeping a reasonably low rejection rate. We choose 15\% after taking into account the size of our dataset and using guidance from~\citet{miller16dimva} to estimate a reasonable labeling capacity. 
}

\revised{
Recall that the calibration constraints are with respect to the calibration set which ideally exhibits minimal drift. It is clear from our evaluation that as concept drift becomes more severe during a deployment, constraints such as those on the rejection rate will be surpassed to some degree. This is the desired outcome---so long as the performance on rejected elements remains low (\ie they would likely be misclassified) we would rather reject drifting examples. 
}

\revised{
\autoref{fig:threshold-tuning} presents an alternative to the optimization used in our previous experiments which is more appropriate if the rejection rate must be kept low. By finding thresholds that minimize the rejection rate on the calibration set with \fone-Score no less than 0.8, during deployment the rejection rate stays much lower, consistently staying below 10\% even as the drift increases. Similar to how the rejection rate begins close to the calibration constraint and then increases in our previous experiments, in this setting the \fone begins close to the calibration constraint, and then decreases. The overall effect here is that the ICE is more conservative in its rejections: while the \fone of kept elements decreases as more incorrect predictions are accepted, the ICE rejects only those predictions that are most likely to be incorrect, keeping the \fone of rejected elements at 0.
}

\revised{
In summary, to estimate how many rejections will be acceptable, we advise practitioners to consider the expected volume of incoming samples, the available resources for processing quarantined examples, and the lifetime of the classifier before being retrained (as drift will likely increase during this period). Next they should identify which metrics are most important, or nonnegotiable, and use these to balance the threshold optimization. As the emergence of concept drift will likely result in the calibration values being surpassed, a ballpark is more important than the exact values.
}

\subsection{Formal Calibration Algorithms}
\label{app:algos}

We present algorithms for our random search calibration and calibration and test procedures for TCEs, ICEs, and CCEs.

\begin{algorithm}[t!]
\caption{Random search threshold calibration}

\label{alg:search}

\DontPrintSemicolon

\footnotesize

\SetKwInOut{Input}{Input}
\Input{y, $\hat{y}$, $pval_c$ }
\KwIn{$\bm{Y} \in \mathcal{Y}^n$, ground truth labels for $n$ examples\newline
	  $\bm{\hat{Y}} \in \mathcal{Y}^n$, predicted labels for $n$ examples\newline
	  $P \in \mathbb{R}^{n \times \miniclasscardinality}$, per-class p-values for $n$ examples}

\SetKwInOut{Parameters}{Parameters}
\Parameters{$m \in \mathbb{R}$, maximum number of iterations\newline
			$\mathcal{F} : \bm{Y} \times \bm{\hat{Y}} \times P \longrightarrow \mathbb{R}$, performance measure to optimize (\eg $F_{1}$)\newline
			$\mathcal{G} : \bm{Y} \times \bm{\hat{Y}} \times P \longrightarrow \mathbb{R}$, performance measure to constrain (\eg kept examples)\newline
			$\mathcal{C} \in \mathbb{R}$, lower bound for constrained measure $\mathcal{G}$}

\KwOut{$\bm{t^{*}}$, a vector of per-class thresholds}
\KwOut{$\bm{t^{*}} \in [0,1]^{\miniclasscardinality}$, a vector of per-class thresholds}

\SetKwData{Counter}{counter}

\BlankLine

$\bm{t^{*}} \leftarrow \bm{0}$ \;
\Counter$ \leftarrow$ 0\;

\While{\Counter < m}{
$\bm{t} \uniformsample [0,1]^{\miniclasscardinality}$
\Comment*{Pick random thresholds}

\uIf{$\mathcal{F}(\bm{Y}, \bm{\hat{Y}}, P; \bm{t}) > \mathcal{F}(\bm{Y}, \bm{\hat{Y}}, P; \bm{t^{*}})$ and $\mathcal{G}(\bm{Y}, \bm{\hat{Y}}, P; \bm{t}) \ge \mathcal{C}$}
{
    $\bm{t^{*}} \leftarrow \bm{t}$\;
}

\uElseIf{$\mathcal{F}(\bm{Y}, \bm{\hat{Y}}, P; \bm{t}) = \mathcal{F}(\bm{Y}, \bm{\hat{Y}}, P; \bm{t^{*}})$ and $\mathcal{G}(\bm{Y}, \bm{\hat{Y}}, P; \bm{t}) > \mathcal{G}(\bm{Y}, \bm{\hat{Y}}, P; \bm{t^{*}})$}
{
    $\bm{t^{*}} \leftarrow \bm{t}$\;
}

\Counter$\leftarrow$ \Counter + 1\;

}

\Return{\bm{$t^{*}$}}\;

\end{algorithm}

\begin{algorithm}[tb]
\caption{Transductive Conformal Evaluator (TCE and \textit{approximate} TCE)}

\label{alg:tce}

\small

\DontPrintSemicolon

\SetKwFunction{Fit}{Fit}
\SetKwFunction{GetThresholds}{Transcend.FindThresholds}
\SetKw{Emit}{emit}
\SetKw{Calibration}{Calibration Phase}
\SetKw{Test}{Test Phase}

\KwIn{$Z = \Lbag z_0, z_1, \dots, z_{n-1} \Rbag$, $n$ training examples\newline
	  $Z^* = \Lbag z^*_0, z^*_1, \dots \Rbag$, stream of test examples\newline
	  $A$, NCM for producing nonconformity scores\newline
	  $k \in \mathbb{N}$, number of folds---TCE is \textit{approximate} when $k < n$}

\KwOut{Stream of boolean decisions $0 = reject, 1 = accept$}

\vskip.5\baselineskip
\nonl \Calibration \;
\algrule

\BlankLine

$\bm{P} \leftarrow \bm{0}$ \;
$i \leftarrow 0$ \;

\textbf{partition} $Z$ equally into $Z^{part} \leftarrow \{\, Z'_0, Z'_1, \dots, Z'_{k-1} \,\}$

\ForEach{partition \ $Z'$ of \ $Z^{part}$}{
	$Z'' \leftarrow Z \setminus Z'$ \;
	$g \leftarrow$ \Fit$(Z'')$ \;
	\ForEach{$z'$ of \ $Z'$}{
	    \Comment*[l]{Predicted label}
		$\hat{y} \leftarrow g(z')$  \;
		\Comment*[l]{Bag of examples with same label}
		$Z'_{\hat{y}} \leftarrow \Lbag z \in Z' : z.y = \hat{y} \Rbag$  \;
		\Comment*[l]{Nonconformity score}
		$\alpha_{z'} \leftarrow A(Z'_{\hat{y}}, z')$  \;
		\Comment*[l]{Nonconformity scores for bag elements}
		$S \leftarrow \Lbag A(Z'_{\hat{y}} \setminus \Lbag z \Rbag) : z \in Z'_{\hat{y}} \Rbag$ \;
		\Comment*[l]{Credibility p-value}
		$p_{z'} \leftarrow \frac{| \alpha \in S : \alpha >= \alpha_{z'} |}{|S|}$  \;
		$\bm{P}_i \leftarrow p_{z'}$ \;
		$i \leftarrow i + 1$ \;
	}
}

$\bm{t}^* \leftarrow $\GetThresholds$(Z, \bm{\hat{Y}}, \bm{P})$ \;

\vskip.5\baselineskip
\nonl \Test \;
\algrule

$g \leftarrow $ \Fit$(Z)$ \;
\ForEach{$z^*$ of \ $Z^*$}{
    \Comment*[l]{Predicted label for test example}
	$\hat{y} \leftarrow g(z^*)$ \;
	\Comment*[l]{Bag of training examples with same label}
	$Z_{\hat{y}} \leftarrow \Lbag z \in Z : z.y = \hat{y} \Rbag$  \;
	\Comment*[l]{Nonconformity score}
	$\alpha_{z^*} \leftarrow A(Z_{\hat{y}}, z^*)$  \;
	\Comment*[l]{Nonconformity scores for bag elements}
	$S \leftarrow \Lbag A(Z_{\hat{y}} \setminus \Lbag z \Rbag) : z \in Z_{\hat{y}} \Rbag$  \;
	\Comment*[l]{Credibility p-value}
	$p_{z^*} \leftarrow \frac{| \alpha \in S \, : \, \alpha >= \alpha_{z^*} |}{|S|}$  \;

	\lIf*{$P_{z^*} < \bm{t}^*_{\hat{y}}$}{\Emit{$0$}  \lElse*{\Emit{$1$}}} \;
}

\end{algorithm}

\begin{algorithm}[tb]
\caption{Inductive Conformal Evaluator (ICE)}
\label{alg:ice}

\small

\DontPrintSemicolon

\SetKwFunction{Fit}{Fit}
\SetKwFunction{GetThresholds}{Transcend.FindThresholds}
\SetKw{Emit}{emit}
\SetKw{Calibration}{Calibration Phase}
\SetKw{Test}{Test Phase}

\KwIn{$Z = \Lbag z_0, z_1, \dots, z_{n-1}\Rbag$, $n$ training examples\newline
	  $Z^* = \Lbag z^*_0, z^*_1, \dots \Rbag$, stream of test examples\newline
	  $A$, NCM for producing nonconformity scores\newline
	  $m$, number of examples to use for calibration}

\KwOut{Stream of boolean decisions $0 = reject, 1 = accept$}

\vskip.5\baselineskip
\nonl \Calibration \;
\algrule

\BlankLine

$\bm{P} \leftarrow \bm{\hat{Y}} \leftarrow \bm{0}$ \;
$i \leftarrow 0$ \;

$Z^{tr} \leftarrow \Lbag z_0, z_1, \dots, z_{n-m-1} \Rbag$ \;
$Z^{cal} \leftarrow \Lbag z_{n-m}, z_{n-m+1}, \dots, z_{n-1} \Rbag $ \;

\ForEach{$z'$ of \ $Z^{cal}$}{
	$g \leftarrow$ \Fit$(Z^{cal} \setminus \Lbag z' \Rbag)$ \;
	\Comment*[l]{Predicted label}
	$\hat{y} \leftarrow \bm{\hat{Y}}_i \leftarrow g(z')$ \;
	\Comment*[l]{Bag of examples with same label}
	$Z^{cal}_{\hat{y}} \leftarrow \Lbag z \in Z^{cal} : z.y = \hat{y} \Rbag$ 	\;
	\Comment*[l]{Nonconformity score}
	$\alpha_{z'} \leftarrow A(Z^{cal}_{\hat{y}}, z')$ \;
	\Comment*[l]{Nonconformity scores for bag elements}
	$S \leftarrow \Lbag A(Z^{cal}_{\hat{y}} \setminus \Lbag z \Rbag) : z \in Z^{cal}_{\hat{y}} \Rbag$ \;
	\Comment*[l]{Credibility p-value} g
	$p_{z'} \leftarrow \frac{| \alpha \in S \, : \, \alpha >= \alpha_{z'} |}{|S|}$ \;
	$\bm{P}_i \leftarrow p_{z'}$ \;
	$i \leftarrow i + 1$ \;
}
$\bm{t}^* \leftarrow $\GetThresholds$(Z, \bm{\hat{Y}}, \bm{P})$ \;

\vskip.5\baselineskip
\nonl \Test \;
\algrule

$g \leftarrow $ \Fit$(Z^{tr})$ \;
\ForEach{$z^*$ of \ $Z^*$}{
	\Comment*[l]{Predicted label for test example}
	$\hat{y} \leftarrow g(z^*)$ \;
	\Comment*[l]{Bag of training examples with same label}
	$Z^{cal}_{\hat{y}} \leftarrow \Lbag z \in Z^{cal} : z.y = \hat{y} \Rbag$ \;
	\Comment*[l]{Nonconformity score}
	$\alpha_{z^*} \leftarrow A(Z^{cal}_{\hat{y}}, z^*)$  \;
	\Comment*[l]{Nonconformity scores for bag elements}
	$S \leftarrow \Lbag A(Z^{cal}_{\hat{y}} \setminus \Lbag z \Rbag) : z \in Z^{cal}_{\hat{y}} \Rbag$  \;
	\Comment*[l]{Credibility p-value}
	$p_{z^*} \leftarrow \frac{| \alpha \in S \, : \, \alpha >= \alpha_{z^*} |}{|S|}$

	\lIf*{$P_{z^*} < \bm{t}^*_{\hat{y}}$}{\Emit{$0$}  \lElse*{\Emit{$1$}}}
}
\end{algorithm}

\begin{algorithm}[tb]
\caption{Cross-Conformal Evaluator (CCE)}
\label{alg:cce}

\small

\DontPrintSemicolon

\SetKwFunction{Fit}{Fit}
\SetKwFunction{GetThresholds}{Transcend.FindThresholds}
\SetKw{Emit}{emit}
\SetKw{Calibration}{Calibration Phase}
\SetKw{Test}{Test Phase}

\KwIn{$Z = \Lbag z_0, z_1, \dots, z_{n-1}\Rbag$, $n$ training examples\newline
	  $Z^* = \Lbag z^*_0, z^*_1, \dots \Rbag$, stream of test examples\newline
	  $A$, NCM for producing nonconformity scores\newline
	  $k \in \{\,2t + 1 : t \in \mathbb{N}\,\}$, number of folds}

\KwOut{Stream of boolean decisions $0 = reject, 1 = accept$}

\vskip.5\baselineskip
\nonl \Calibration \;
\algrule

\BlankLine

$\bm{P} \leftarrow \bm{\hat{Y}} \leftarrow \bm{G} \leftarrow \bm{t^{*}} \leftarrow \bm{0}$ \;
$i \leftarrow j \leftarrow 0$ \;

\textbf{partition} $Z$ equally into $\{\, Z'_0, Z'_1, \dots, Z'_{k-1} \,\}$

\ForEach{$j$ of \ $\{\, 0,1,\dots,k-1 \,\}$}{

\ForEach{$z'$ of \ $Z'_j$}{
	$g \leftarrow$ \Fit$(Z'_j \setminus \Lbag z' \Rbag)$ \;
	\Comment*[l]{Predicted label}
	$\hat{y} \leftarrow \bm{\hat{Y}}_{j,i} \leftarrow g(z')$  \;
	\Comment*[l]{Bag of examples with same label}
	$Z'_{j_{\hat{y}}} \leftarrow \Lbag z \in Z'_j : z.y = \hat{y} \Rbag$  \;
	\Comment*[l]{Nonconformity score}
	$\alpha_{z'} \leftarrow A(Z'_{j_{\hat{y}}}, z')$  \;
	\Comment*[l]{Nonconformity scores for bag elements}
	$S \leftarrow \Lbag A(Z'_{j_{\hat{y}}} \setminus \Lbag z \Rbag) : z \in Z'_{j_{\hat{y}}} \Rbag$  \;
	\Comment*[l]{Credibility p-value}
	$\bm{P}_{j,i} \leftarrow \frac{| \alpha \in S \, : \, \alpha >= \alpha_{z'} |}{|S|}$  \;
	$i \leftarrow i + 1$ \;
}

$\bm{G}_j \leftarrow $ \Fit$(Z \setminus Z'_j)$ \;
$\bm{T}^*_j \leftarrow $\GetThresholds$(Z'_j, \bm{\hat{Y}_j}, \bm{P}_j)$ \;
}

\vskip.5\baselineskip
\nonl \Test \;
\algrule
$s \leftarrow 0$ \;
\ForEach{$z^*$ of \ $Z^*$}{
	\ForEach{$j$ of \ $\{\, 0,1,\dots,k-1 \,\}$} {
	\Comment*[l]{Predicted label for test example}
	$\hat{y} \leftarrow G_j(z^*)$  \;
	\Comment*[l]{Bag of training examples with same label}
	$Z'_{j_{\hat{y}}} \leftarrow \Lbag z \in Z'_j : z.y = \hat{y} \Rbag$  \;
	\Comment*[l]{Nonconformity score}
	$\alpha_{z^*} \leftarrow A(Z'_{j_{\hat{y}}}, z^*)$  \;
	\Comment*[l]{Nonconformity scores for bag elements}
	$S \leftarrow \Lbag A(Z'_{j_{\hat{y}}} \setminus \Lbag z \Rbag) : z \in Z'_{j_{\hat{y}}} \Rbag$  \;
	\Comment*[l]{Credibility p-value}
	$p_{z^*} \leftarrow \frac{| \alpha \in S \, : \, \alpha >= \alpha_{z^*} |}{|S|}$  \;
	\Comment*[l]{Track positive evaluations}
	\lIf*{$P_{z^*} \geq \bm{T}^*_{j_{\hat{y}}}$}{$s \leftarrow s + 1$}  \;
	}
	\Comment*[l]{Majority vote for final decision}
	\lIf*{$s < k / 2$}{\Emit{$0$}  \lElse*{\Emit{$1$}}}  \;
}
\end{algorithm}

\end{appendix}

\end{document}